\documentclass[]{interact}
\usepackage{epstopdf}
\usepackage[caption=false]{subfig}
\usepackage[doublespacing]{setspace}
\usepackage[numbers,sort&compress]{natbib}
\bibpunct[, ]{[}{]}{,}{n}{,}{,}
\makeatletter
\def\NAT@def@citea{\def\@citea{\NAT@separator}}
\makeatother

\theoremstyle{plain}

\theoremstyle{definition}

\theoremstyle{remark}

\usepackage{color}

\begin{document}

\articletype{ARTICLE TEMPLATE}

\title{Energy Dependent Calculations of Fission Product, Prompt, and Delayed Neutron Yields for Neutron Induced Fission on $^{235}$U, $^{238}$U, and $^{239}$Pu}

\author{
  \name{S. Okumura\textsuperscript{a}\thanks{CONTACT S. Okumura. Email: s.okumura@iaea.org},
        T. Kawano\textsuperscript{b},
        A.E. Lovell\textsuperscript{b},
        and T. Yoshida\textsuperscript{c}}
  \affil{\textsuperscript{a}NAPC--Nuclear Data Section, International Atomic Energy Agency, Vienna A-1400, Austria;\\
         \textsuperscript{b}Theoretical Division, Los Alamos National Laboratory, Los Alamos, NM 87545, USA;\\
         \textsuperscript{c}Laboratory for Advanced Nuclear Energy, Tokyo Institute of Technology, Tokyo 152-8550, Japan}}

\maketitle

\begin{abstract}
  We perform energy dependent calculations of independent and cumulative
fission product yields for $^{235}$U, $^{238}$U, and $^{239}$Pu in the
first chance fission region. Starting with the primary fission
fragment distributions taken from available experimental data and
analytical functions based on assumptions for the excitation energy
and spin-parity distributions, the Hauser-Feshbach statistical decay
treatment for fission fragment de-excitation is applied to more than
1,000 fission fragments for the incident neutron energies up to
5~MeV. The calculated independent fission product yields are then used
as an input of $\beta$-decay to produce the cumulative yield, and
summation calculations are performed. Model parameters in these
procedures are adjusted by applying the Bayesian technique at the
thermal energy for $^{235}$U and $^{239}$Pu and in the fast energy
range for $^{238}$U. The calculated fission observable quantities,
such as the energy-dependent fission yields, and prompt and delayed
neutron yields, are compared with available experimental data. We also
study a possible impact of the second chance fission opening on the
energy dependence of the delayed neutron yield by extrapolating the
calculation.

\end{abstract}


\def\nup{\overline{\nu}_p}
\def\nud{\overline{\nu}_d}

\section{Introduction}
Nuclear systems that involve the nuclear fission process often require
very high accuracy of both prompt and delayed neutron multiplicity
data, $\nup$ and $\nud$, albeit model predictions for these quantities
are not yet at the satisfactory level. For major fissioning systems,
such as the neutron-induced reaction on $^{235}$U, more than 99\%
neutrons are the prompt fission neutrons, which are emitted from
highly excited two fission fragments formed just after
fission. Typically there are more than 1000 fission fragments, and
2--3 prompt neutrons per fission are emitted.  Whereas a small
fraction ($\sim 1$\%) in the total neutron yield is produced during
the $\beta$-decay chain of fission products, and approximately 270
nuclides have been identified as precursors for the delayed neutron
emission~\cite{Brady1989,LA-11534-T}. Ideally we can calculate $\nup$
and $\nud$ by summing up all the decaying compound nuclei weighted by
the fission yields, which is the so-called summation calculation
({\it e.g.} Ref.~\cite{Yoshida1999}). This method, however, requires a lot
of well-tuned model inputs. This was partly done in our previous
study~\cite{Okumura2018} for the prompt neutron emission.

Since the discovery of delayed neutron by Roberts {\it et al.} shortly
after the discovery of nuclear fission in 1939~\cite{Roberts1939},
despite its tiny fraction, the delayed neutron has attracted
people in various scientific communities, as its quite important
role in keeping the thermal reactors critical, as well as the reactor
systems containing high burn-up fuel, and transmutation of minor
actinides. The delayed neutron yield has been 
measured~\cite{Keepin1957,Masters1969,Krick1972,Piksaikin2002} and
repeatedly evaluated~\cite{Evans1973,Tuttle1975,Brady1989,Yoshida2002}
for various fissioning systems at several incident neutron energies.
Some models for predicting the time-dependent delayed neutron yield
have been proposed~\cite{Sikora1983,Lendel1986,Waldo1981}, and these
studies pointed out the importance of fission yield data to perform
these model calculations.

When an incident neutron energy goes higher, it is natural that $\nup$
also increases monotonously, since the formed compound nucleus has
larger available total energy. However, in contrast to $\nup$, $\nud$
shows totally different behavior, depending on how the delayed neutron
precursors are produced. There still exists challenges to understand
peculiar energy-dependence of $\nud$, {\it i.e.,} a slight increase in
the yield from thermal to 3~MeV and a steep decrease above 4 MeV as
seen in $^{235,238}$U. To account for the abrupt changes, the
evaluated $\nud$ data in nuclear data libraries, JENDL-4.0~\cite{JENDL4}
and ENDF/B-VIII~\cite{ENDF8}, include a very crude piecewise linear
function to represent experimental data.

Such the energy-dependent behavior has not yet been explained
theoretically. Alexander {\it et al.}~\cite{Alexander1977} first
interpreted the energy-dependence in $\nud$ by taking into account the
odd-even effect of fission products. Ohsawa {\it et
al.}~\cite{Ohsawa2002,Ohsawa2007} introduced the multimodal random
neck-rupture model~\cite{Brosa1990} and fission mode
fluctuations~\cite{Hambsch1989} to explain the
energy-dependence. Minato~\cite{Minato2018} proposed a model to
reproduce the energy-dependence of $\nud$ based on the fission yield
using Katakura's systematics~\cite{JAER-_Research-2003-004}.
Although an explicit statistical decay was not performed in Minato's
model —-- hence the calculated $\nup$ and $\nud$ are independent of one
another —-- it also supports Alexander's observation: the odd-even
effect in the the charge distribution is important.
Recently the odd-even effect was
explained by applying the microscopic number projection
method~\cite{Verriere2019}.


By extending the Hauser-Feshbach Fission Fragment Decay (HF$^3$D)
model~\cite{Okumura2018} to the $\beta$-decay process, consistency
among the independent and cumulative fission yields $Y_I(Z,A)$ and
$Y_C(Z,A)$, and neutron multiplicities $\nup$ and $\nud$, is
automatically guaranteed. In this model, we start with the fission
fragment distribution $Y(Z,A,E_{ex},J,\Pi)$ characterized by the
distributions of mass and charge, excitation energy, and
spin/parity. We perform the Hauser-Feshbach statistical decay for the
excited fission fragments to calculate the independent fission yields
$Y_I(Z,A)$ and $\nup$. A successive $\beta$-decay calculation gives
the cumulative yields $Y_C(Z,A)$ and $\nud$.  The model parameters are
adjusted to reproduce experimental data at thermal by applying the
Bayesian technique, and we extrapolate the calculation to the second
chance fission threshold. In this paper, we limit ourselves mainly to
first-chance fission, because more uncertain parameters will be
involved in the multi-chance fission case. Although we study the
multi-chance fission case elsewhere~\cite{Lovell2021}, here, we
briefly explore a possible impact of the second-chance opening with a
particular focus on $\nud$.

\section{Methods}
\subsection{Hauser-Feshbach statistical decay and $\beta$-decay calculations}
\subsubsection{Sources of energy-dependence}

The energy dependence of the independent and cumulative fission product
yields (FPY) arises from properties of some model parameters. The
primary fission fragment distribution $Y_P(Z,A)$, often approximated
by a few Gaussian forms, gradually changes the shape as the incident
neutron energy increases.  When the excitation energy of the fissioning compound
system increases, the fission path after the second barrier spreads
along the most probable path, hence the asymmetric terms will have
wider width, and the peaks of distributions will be lower to satisfy
the normalization condition.

The energy dependence of total kinetic energy (TKE) is also the one of
the related physical observables of predicting energy-dependent
FPY. We often see that the experimental data of TKE decrease
monotonously for some major fissioning nuclides such as $^{235,238}$U
and $^{239}$Pu~\cite{ANLNDM64,Madland2006,Zoeller1995}, except at very low
energies~\cite{Duke2014,LA-UR-15-28829}.

The anisothermal parameter $R_T$, which changes the number of prompt
neutrons removed from the fission fragments, often needs to be larger
than unity to reproduce the neutron multiplicity distribution as a
function of fragment mass number, $\nu(A)$. The reason of this is
still unclear. It might be natural to assume $R_T=1$ by the
phase-space argument, where the total excitation energy would be
shared by the two fragments according to the number of available
states.  The odd-even effect in the charge distribution of Wahl's
$Z_p$ model~\cite{LA13928,Wahl1988} might decrease at higher
excitation energies, where a particular nuclear structure effect no
longer persists. Since the original Whal systematics does not 
consider any energy dependence of the odd-even effect, we 
incorporate the energy dependencies of these parameters, 
yet phenomenological parameterization is applied.

\subsubsection{Generation of the fission fragment distribution}

The primary fission fragment distributions are the key ingredient in
the prompt neutron emission calculation. While this is a complicated
multi-dimensional distribution, including energy, spin, parity, etc.,
we demonstrated that the numerical integration over all these
distributions is feasible by the Hauser-Feshbach Fission Fragment
Decay (HF$^3$D) model.  The model produces various fission observables
simultaneously, {\it e.g.}, the prompt neutron multiplicity
$\overline{\nu}_p$, independent FPY $Y_I(Z,A)$, and isomeric ratio
(IR)~\cite{Okumura2018}. Since the method and relevant
equations are explained elsewhere~\cite{Okumura2018}, a brief
description as well as newly developed components will be given here.

The primary fission fragment yield $Y_P(Z,A)$ is constructed by five
(or seven if needed) Gaussians fitted to experimental primary fission fragment
mass distributions of neutron induced reaction on $^{235}$U,
$^{238}$U, and $^{239}$Pu. A charge distribution for a given mass
number is generated by the $Z_p$ model~\cite{Wahl1988} of Wahl's
systematics~\cite{LA13928} implemented in the HF$^3$D model.

TKE as a function of primary fission fragment mass ${\rm TKE}(A)$ is
also generated based on the experimental data, which yields the
average excitation energy of each fragment.  An $A$-average of ${\rm
TKE}(A)$ gives a TKE value at a given neutron incident energy
${\rm TKE}(E)$, and the variance of ${\rm TKE}(A)$ gives the excitation
energy distribution. By combining with the distributions of excitation
energy $E_{ex}$, spin $J$, and parity $\Pi$ described in the previous
work~\cite{Okumura2018}, an initial configuration of fission fragment
compound nucleus $Y_P(Z,A,E_{ex},J,\Pi)$ is fully characterized. The
Hauser-Feshbach theory is applied to the statistical decay of
generated $Y_P(Z,A,E_{ex},J,\Pi)$.  The experimental data sets used in
this study are listed in Tables~\ref{tbl:Y_A}, \ref{tbl:TKE_A},
and \ref{tbl:TKE_E}.

The functional forms for ${\rm TKE}(A)$ and ${\rm TKE}(E)$ are given 
in our former work~\cite{Okumura2018}, and the parameters of
these functions for $^{235}$U are the same as before. Those for $^{239}$Pu
were taken from the CGMF code~\cite{Talou2018}. Because there is no
primary fission fragment data for $^{238}$U at thermal, the parameters in
$Y_P(Z,A)$ and ${\rm TKE}(A)$ are determined in the 1.1 -- 1.3~MeV region.
The obtained $Y_P(Z,A)$ is given later, and ${\rm TKE}(A)$ is
\begin{equation}
  {\rm TKE}(A_h)
      = (348.371 - 1.274 A_h)
      \left\{
        1 - 0.1800 \exp\left(-\frac{(A_h - A_m)^2}{59.199}\right)
      \right\} \quad\mbox{MeV}\ ,
\end{equation}
and ${\rm TKE}(E)$ is
\begin{equation}
  {\rm TKE}(E)
      = 171.11 -0.320 E_n \quad\mbox{MeV}\ ,
\end{equation}
where the incident energy $E_n$ is in MeV.

\begin{table}[htb]
  \centering
  \caption{Experimental data of mass distributions included in the parameter
     fitting of $Y_P(A,E)$.}
  \label{tbl:Y_A}
\begin{tabular}{ccll}
  \hline
 Nuclide  & Energy (MeV) & Author \& Reference \\
  \hline
$^{235}$U  & $2.53\times10^{-8}$      & Baba {\it et al.}         & \cite{Baba1997} \\
          & $2.53\times10^{-8}$      & Hambsch                   & \cite{Hambsch} \\
          & $2.53\times10^{-8}$      & Pleasonton {\it et al.}   & \cite{Pleasonton1972} \\
          & $2.53\times10^{-8}$      & Simon {\it et al.}        & \cite{Simon1990} \\
          & $2.53\times10^{-8}$      & Straede {\it et al.}      & \cite{Straede1987} \\
          & $2.53\times10^{-8}$      & Zeynalov {\it et al.}     & \cite{Zeynalov2006} \\
          & $2.53\times10^{-8}$ -- 7 & D'yachenko {\it et al.}   & \cite{Dyachenko1969} \\
  \hline
$^{238}$U  & 1.11, 1.25   & Goverdovskiy {\it et al.} & \cite{Goverdovskiy2000} \\
          & 1.2 -- 5.8 & Vives {\it et al.}    & \cite{Vives2000} \\
  \hline
$^{239}$Pu & $2.53\times10^{-8}$ -- 4.48 & Akimov {\it et al.}    & \cite{Akimov1971} \\
          & $2.53\times10^{-8}$      & Surin {\it et al.}        & \cite{Surin1972} \\
          & $2.53\times10^{-8}$      & Wagemans {\it et al.}     & \cite{Wagemans1984} \\
          & $2.53\times10^{-8}$      & Schillebeeckx {\it et al.}& \cite{Schillebeeckx1992} \\
          & $2.53\times10^{-8}$      & Nishio {\it et al.}       & \cite{Nishio1995} \\
          & $2.53\times10^{-8}$      & Tsuchiya {\it et al.}     & \cite{Tsuchiya2000} \\
  \hline
\end{tabular}
\end{table}

\begin{table}[htb]
  \centering
  \caption{Experimental data included in the parameter fitting of ${\rm TKE}(A)$.}
\label{tbl:TKE_A}
\begin{tabular}{ccll}
  \hline
 Nuclide & Energy (MeV) & Author \& Reference \\
  \hline
$^{235}$U & $2.53\times10^{-8}$      & Baba {\it et al.}       & \cite{Baba1997} \\
         & $2.53\times10^{-8}$      & Hambsch                 & \cite{Hambsch} \\
         & $2.53\times10^{-8}$      & Simon {\it et al.}      & \cite{Simon1990} \\
         & $2.53\times10^{-8}$      & Zeynalov {\it et al.}   & \cite{Zeynalov2006} \\
         & $2.53\times10^{-8}$      & D'yachenko {\it et al.} & \cite{Dyachenko1969} \\
  \hline
$^{238}$U  & 1.2   & Vives {\it et al.}  & {\cite{Vives2000}} \\
  \hline
$^{239}$Pu & $2.53\times10^{-8}$      &  Surin {\it et al.}    & \cite{Surin1972} \\
          & $2.53\times10^{-8}$      & Wagemans {\it et al.}  & \cite{Wagemans1984} \\
          & $2.53\times10^{-8}$      & Nishio {\it et al.}    & \cite{Nishio1995} \\
          & $2.53\times10^{-8}$      & Tsuchiya {\it et al.}  & \cite{Tsuchiya2000} \\
  \hline
\end{tabular}
\end{table}

\begin{table}[htb]
  \centering
  \caption{Experimental data included in the parameter fitting of ${\rm TKE}(E)$.}
\label{tbl:TKE_E}
\begin{tabular}{ccll}
  \hline
 Nuclide & Energy (MeV) & Author \& Reference \\
  \hline
$^{235}$U & 0.18 -- 8.83 & Meadows and Budtz-J{\o}rgensen & \cite{ANLNDM64} \\
         & $2.53\times10^{-8}$ -- 35.5    & Duke                        & \cite{Duke2014} \\
  \hline
$^{238}$U & 1.5 -- 400.0     & Z{\"o}ller {\it et al.}     & \cite{Zoeller1995} \\
         & 1.4 -- 28.3      & Duke {\it et al.}        & \cite{Duke2016} \\
  \hline
$^{239}$Pu & 0.05 -- 5.3 & Akimov {\it et al.}     & \cite{Akimov1971} \\
          & $2.53\times10^{-8}$ -- 3.55 & Vorobeva {\it et al.}     & \cite{Vorobeva1974} \\
          & 0.5 -- 50 & Meierbachtol {\it et al.} & \cite{Meierbachtol2016}\\
  \hline
\end{tabular}
\end{table}

\subsubsection{Model Parameters}

The Gaussian terms for $Y_P(A)$ are parameterized as
\begin{equation}
 Y_P(A) = \sum_{i=1}^5 \frac{F_i}{\sqrt{2\pi}\sigma_i}
        \exp\left\{
              -\frac{(A-A_m + \Delta_i)^2}{2\sigma_i^2}
            \right\} \ ,
\label{eq:fivegaussian}
\end{equation}
where $\sigma_i$ and $\Delta_i$ are the Gaussian parameters, the index
$i$ runs from the low mass side, and the component of $i=3$ is for the
symmetric distribution ($\Delta_3 = 0$). $A_m=A_{\rm CN}/2$ is the
mid-point of the mass distribution, $A_{\rm CN}$ is the mass number of
fissioning compound nucleus, and $F_i$ is the fraction of each
Gaussian component. The symmetric shape of $Y_P(A)$ ensures implict
relations of $F_1 = F_5$, $F_2 = F_4$, etc.

We assume that the energy sharing between the complementary light and
heavy fragments is followed by the anisothermal
model~\cite{Ohsawa1999,Ohsawa2000}, which is defined by the ratio of
effective temperature $T_L$ and $T_H$ in the light and
heavy fission fragments,
\begin{equation}
  R_T = \frac{T_L}{T_H} = \sqrt{\frac{a_H U_L}{a_L U_H}} \ ,
\end{equation}
where $U$ is the excitation energy corrected by the pairing
energy~\cite{Gilbert1965}, and $a$ is the level density parameter
including the shell correction energy.

There are several estimates of $R_T$ for different fissioning systems.
In the case of thermal neutron induced fission on $^{235}$U, a
constant $R_T$ reasonably reproduces the experimental $\nu(A)$
data~\cite{Okumura2018}, and Talou {\it et
al.}~\cite{Talou2008,Talou2009} showed the cases of
$^{239}$Pu(n$_{th}$,f), and $^{252}$Cf spontaneous fission.  However,
it has been reported that better reproduction of experimental data is
achieved by mass-dependent $R_T$
parameters~\cite{Litaize2010,Manailescu2011,Talou2011,Becker2013}.  In
the present work, we do not explore all possible functional forms of
$R_T$.  Instead, a simple energy-dependence is introduced as
\begin{equation}
  R_T = 
  \left\{
  \begin{array}{lr}
     R_{T0} + E_n R_{T1}\ , & R_{T0} + E_n R_{T1} \ge 1 \\
     1                 \ , & \mbox{otherwise} \\
  \end{array}
  \right. \ ,
  \label{eq:RT}
\end{equation}
where $R_{T0}$ and $R_{T1}$ are model parameters. As we expect $R_T$
decreases as the incident energy, $R_{T1} < 0$.

In Wahl's $Z_p$ model the even-odd effect in the $Z$-distribution is given as
\begin{equation}
f = \left\{
    \begin{array}{ccc}
      F_Z F_N     &  Z \mbox{even} & N \mbox{even} \\
      F_Z / F_N   &  Z \mbox{even} & N \mbox{odd} \\
      F_N / F_Z   &  Z \mbox{odd} & N \mbox{even} \\
      1/(F_Z F_N) &  Z \mbox{odd} & N \mbox{odd} \\
    \end{array}\right. \ ,
\end{equation}
where $F_Z \ge 1$ and $F_N \ge 1$ are parameterized and tabulated by
Wahl.  This equation gives higher yields when $Z$ and/or $N$ are the
even number. We expect such even-odd staggering will be mitigated when
a fissioning system has higher excitation energy. We model the
reduction in the even-odd effect by
\begin{eqnarray}
  F_Z &=& 1.0 + (F_Z^{\rm W} - 1.0)  f_Z \ , 
  \label{eq:FZ} \\
  F_N &=& 1.0 + (F_N^{\rm W} - 1.0)  f_N \ ,
  \label{eq:FN}
\end{eqnarray}
where $F_Z^{\rm W}$ and $F_N^{\rm W}$ are the parameters in Wahl's
systematics, and $f_Z$ and $f_N$ are the scaling factor as inputs.
These scaling factors are also linear functions of incident neutron
energy, $f_i = f_{i0} + E_n f_{i1}, \quad i = Z,N$.

\subsection{$\beta$-decay calculation}
The HF$^3$D model produces the independent fission product yields
$Y_I(Z,A)$, as well as the meta-stable state production when the
nuclear structure data indicate that the level half-life is long
enough (typically more than 1~ms.)  Here we add a meta-state index $M$
to specify the isomers explicitly, $Y_I(Z,A,M)$ and $Y_C(Z,A,M)$.  The
cumulative yields are calculated in a time-independent manner, hence
$Y_I(Z,A,M)$ and $Y_C(Z,A,M)$ are simply connected by the decay
branching ratios~\cite{Kawano2013b}. The decay data included are the
half-lives $T_{1/2}$, the decay mode ($\alpha$-decay, $\beta^-$-decay,
delayed neutron emission, {\it etc.}), and the branching ratios to
each decay mode. They are taken from ENDF/B-VIII decay data
library. We also considered JENDL-4.0 decay data library, however the
result is not so different.

When a decay branch includes a neutron emission mode, this nuclide
is identified as a $\beta$-delayed neutron precursor.  The delayed
neutron yield from this $i$-th precursor is calculated as $\nu_d(i) = Y_C(i)
b_i N_d$, where $b_i$ is the branching ratio to the neutron-decay mode,
and $N_d$ is usually one unless multiple neutron emission is allowed.
The total delayed neutron yield $\overline{\nu}_d$ is $\sum_i \nu_d(i)$.

\subsection{Adjustment of Hauser-Feshbach model calculation parameters by Bayesian technique}
An optimization procedure of the HF$^3$D model parameters is a
non-linear multi-dimensional least-squares problem. Albeit such
complex problem might be solved by the modern technology, this will be
a hefty computation and beyond our scope. Instead, we perform a
relatively small-scale adjustment of the model parameters to reproduce
some of the fission product yield data by applying the Bayesian
technique with the KALMAN code~\cite{Kawano2000}.  The model
parameters are first estimated by comparing with the most sensitive
quantities. They are our prior. Then the prior parameters are
adjusted simultaneously by fitting to the experimental data. Although
it is always ideal to use raw experimental data, we use the
evaluated values that should be representative of available
experimental data.  However, it should be noted that we are not tying
to reproduce the evaluation, but to find a consistent solution among
different fission observable.

The model parameters to be included in the KALMAN calculation are the
first and second Gaussian parameters (fraction $F_i$, width
$\sigma_i$, and mass shift $\Delta_i$ for $i = 1$ and 2.) We fix the
symmetric Gaussian, because it does not have any sensitivities to the
experimental data included in this study, and its fraction is too
small anyway. We also include the anisothermal $R_T$ parameter, the
spin factor $f_J$, and the scaling factor in Eqs.~(\ref{eq:FZ}) and
(\ref{eq:FN}). The adjustment is performed at the thermal
energy (or at relatively low energy for $^{238}$U), and the
energy-dependent parts in these model parameters are fixed.

The sensitivity matrix $C$ is defined as
\begin{equation}
  c_{ij} = \frac{\partial d_i}{\partial p_j} \ , \quad
  1 \le i \le N \ , \quad
  1 \le j \le M \ ,
\end{equation}
where $P = (p_1, p_2, \ldots)$ is the model parameter vector, and $D =
(d_1, d_2, \ldots)$ is the data vector containing the calculated
values. The partial derivatives are calculated numerically.  The
KALMAN code linearizes the model calculation as
\begin{equation}
  D = F(P) \simeq F(P_0) + C \left( P - P_0 \right) \ ,
 \label{eq:linear}
\end{equation}
where $F(P)$ stands for a model calculation with a given parameter $P$,
and $P_0$ is the prior parameter vector.

It is not so easy to impose a constraint $2F_1 + 2F_2 + F_3 = 2$ on
the Gaussian fractions during the adjustment process, {\it e.g.}, when
the $F_1$ parameter is perturbed as $F_1+\delta$, the sum exceeds 2;
$2(F_1+\delta) + 2F_2 + F_3 = 2+2\delta$.  However, we renormalize the
fractions internally
\begin{equation}
  F_j' = F_j - \frac{2\delta}{2 + 2\delta} F_j 
       = F_j\left( 1 - \frac{\delta}{1+\delta} \right) 
\end{equation}
to assure the sum to be 2. $F_j'$ is the actual fraction inside the calculations,
and $F_j$ is not necessarily normalized but represents a model input.

\section{Results}
\subsection{Adjusted model parameters at thermal energy}
\subsubsection{Parameter adjustment for $^{235,238}$U and $^{239}$Pu}

The prior Gaussian parameters, $R_{T0}$, $f_J$, and TKE for $^{235}$U
at the thermal energy are taken from our previous
study~\cite{Okumura2018}. When we modify TKE, ${\rm TKE}(A)$ is
automatically shifted to make sure the $A$-average coincides with the given
TKE value. The original Walh's $Z_p$ model is also employed as the
prior parameter, which means $f_{Z0} = f_{N0} = 1$.  They are shown in
the second column of Table~\ref{tbl:parameter235}. These parameters
are adjusted to reproduce the cumulative fission product yields of
$^{95}$Zr, $^{97}$Zr, $^{99}$Mo, $^{132}$Te, $^{140}$Ba, and
$^{147}$Nd at thermal, as well as $\nup$ and $\nud$.  Now we have 11
parameters $(M=11)$ and 8 data ($N=8$.)

\begin{figure}
  \begin{center}
    \resizebox{0.48\textwidth}{!}{\includegraphics{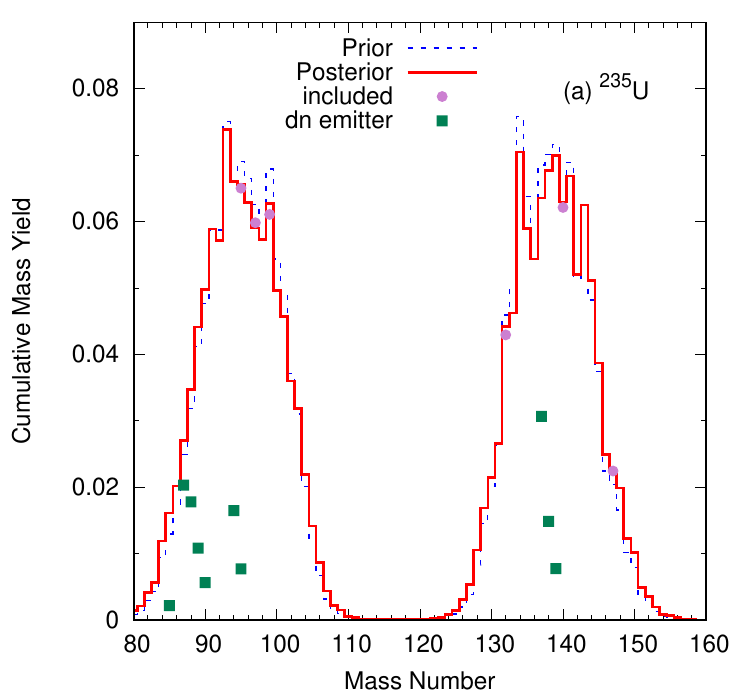}} \\
    \resizebox{0.48\textwidth}{!}{\includegraphics{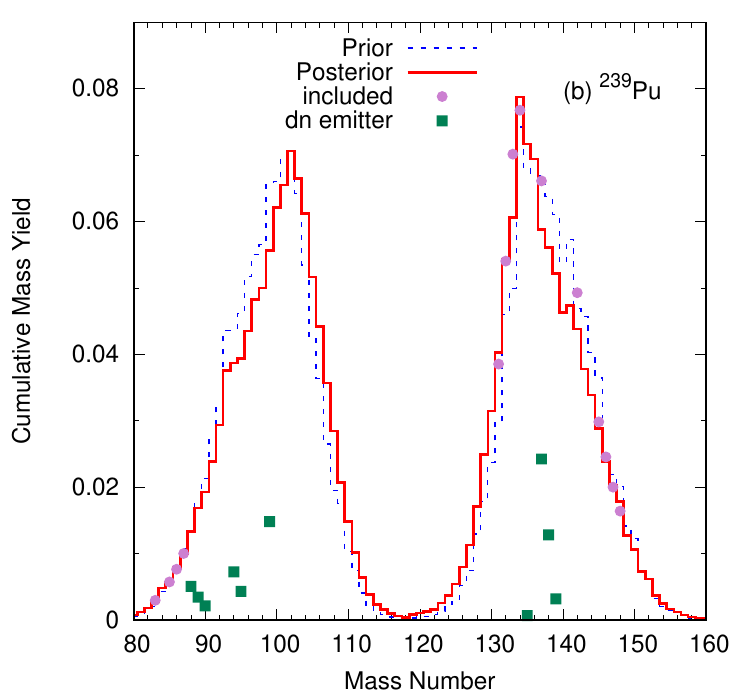}} \\
    \resizebox{0.48\textwidth}{!}{\includegraphics{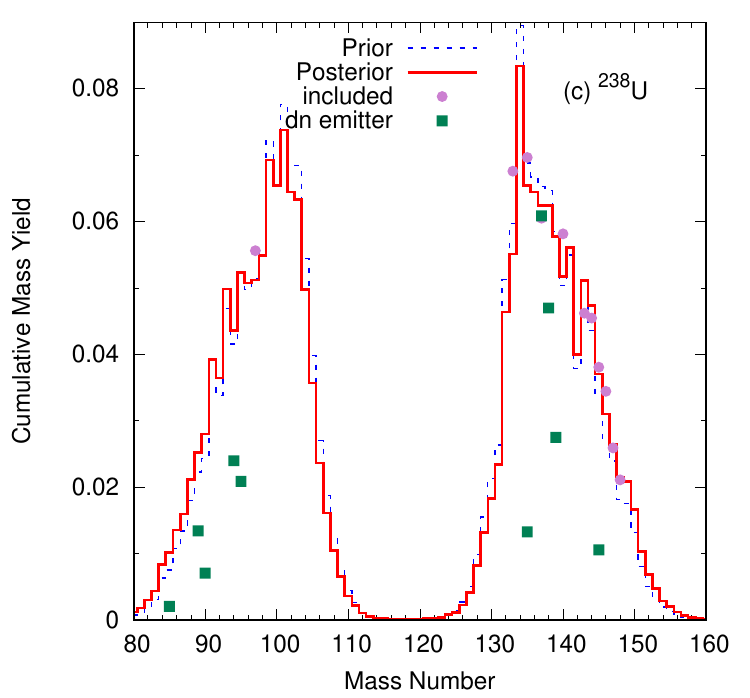}}
  \end{center}

  \caption{Calculated mass chain yields with the prior and posterior
    model parameters for $^{235}$U, $^{239}$Pu, and $^{238}$U. They
    are compared with some selected cumulative fission product yield
    data by the circles.  The squares are the mass chains that include
    major delayed neutron emitters.}

  \label{fig:ya}
\end{figure}

With the prior parameters, the calculated $\nup$ of 2.38 is slightly
lower than the evaluated values of 2.41 (ENDF/B-VIII) and 2.42
(JENDL-4.0), while the prior $\nud$ of 0.0195 is 23\% larger than the
value found in both libraries, 0.0159.  The adjustment reconciles
these discrepancies with the better known values, and the posterior
parameters yield $\nup=2.415$ and $\nud=0.0169$.  The posterior
parameters with their uncertainties and correlation matrix are given
in Table~\ref{tbl:parameter235}.  Since the actual changes in $Y_P(A)$
are very modest, and the posterior parameters equally reproduce the
experimental data of mass distribution, we do not include the
comparison plot here. Figure~\ref{fig:ya} (a) is the mass chain yield
with the prior and posterior parameters. The ENDF evaluated values are
also compared. This figure also shows some mass-chains that contain
major $\beta$-delayed neutron emitters. The reduction in $\nud$ is, in
part, caused by the smaller posterior yields of $A=137$ and 94, which
include $^{137}$I and $^{94}$Rb. While these masses were not included
in the adjustment, the sensitivity of $\nud$ to these masses
implicitly demands the reduction of these mass-chains.

When the prior $R_T$ and $f_J$ parameters are determined, we compare
the neutron multiplicity distribution $P(\nu)$ with the experimental
data. The posterior parameters modify the calculated $P(\nu)$ but not
so significantly. The calculated $P(\nu)$ still agrees fairly well with the
data.

\begin{table}

  \caption{Prior and posterior model parameters defined in
  Eqs.~(\ref{eq:fivegaussian}), (\ref{eq:RT}), (\ref{eq:FZ}),
  and (\ref{eq:FN}), as well as the spin saling factor $f_J$. These
  parameters are dimensionless quantities, except TKE is in MeV.}

  \label{tbl:parameter235}
  \begin{center}
    \begin{tabular}{crrrrrrrrrrrrrr}
\hline
           & pri    & post & \multicolumn{12}{c}{uncertainty[\%] and correlation [\%]}  \\
\hline
$F_1$     &   0.793 &   0.824 &   4.3&$ 100$&      &      &      &      &      &      &      &      &      &      \\
$\sigma_1$&   4.83 &   5.05 &   1.4&$  41$&$ 100$&      &      &      &      &      &      &      &      &      \\
$\Delta_1$&   23.0 &    23.1 &   0.5&$ -36$&$ -56$&$ 100$&      &      &      &      &      &      &      &      \\
$F_2$     &   0.205 &   0.197 &   4.7&$  22$&$ -40$&$  36$&$ 100$&      &      &      &      &      &      &      \\
$\sigma_2$&   2.73 &   2.92 &   3.1&$ -33$&$   1$&$  28$&$  34$&$ 100$&      &      &      &      &      &      \\
$\Delta_2$&    15.6 &    15.2 &   0.7&$  -1$&$  14$&$  39$&$   0$&$  11$&$ 100$&      &      &      &      &      \\
$f_{Z0}$  &   1.00 &   1.78 &   6.6&$   0$&$  -7$&$  48$&$   0$&$   1$&$  41$&$ 100$&      &      &      &      \\
$f_{N0}$  &   1.00 &   0.97 &  20.6&$   0$&$   2$&$   0$&$   0$&$   3$&$  -1$&$   2$&$ 100$&      &      &      \\
$R_{T0}$  &   1.20 &   1.29 &   3.8&$  -3$&$ -10$&$ -64$&$   3$&$ -13$&$ -49$&$ -51$&$   0$&$ 100$&      &      \\
$f_J$     &   3.00 &   2.96 &   4.9&$   6$&$ -30$&$  16$&$  -6$&$  -9$&$ -23$&$   7$&$   0$&$   0$&$ 100$&      \\
TKE       &   170.5 &   170.1 &   0.1&$  -7$&$  25$&$ -27$&$   6$&$  11$&$   1$&$  -5$&$  -1$&$  13$&$ -82$&$ 100$\\
\hline
    \end{tabular}
  \end{center}
\end{table}

\subsubsection{Parameter adjustment for $^{239}$Pu}

The Gaussian parameters obtained by fitting to the experimental $Y_P(A)$
for $^{239}$Pu are
\begin{eqnarray}
   \Delta_1 &=& -\Delta_5 = 14.09 + 0.0994 E_n \ ,\\
   \Delta_2 &=& -\Delta_4 = 20.08 + 0.2940 E_n \ ,\\
   \sigma_1 &=& \sigma_5 =   3.26 + 0.2000 E_n \ ,\\
   \sigma_2 &=& \sigma_4 =   6.58 + 0.1969 E_n \ ,\\
   \sigma_3 &=& 10.0 \ ,
\end{eqnarray}
where $E_n$ in MeV. The fractions of each Gaussian are given by
\begin{eqnarray}
   F_1 &=& F_5 = 0.2483 + 0.0074 E_n \ , \label{eq:FjPu239a}\\
   F_2 &=& F_4 = 0.7184 - 0.0075 E_n \ , \label{eq:FjPu239b}\\
   F_3 &=& 0.003 + 0.003 E_n \ .          \label{eq:FjPu239c}
\end{eqnarray}

The adjustment procedure for $^{239}$Pu at thermal includes the same
parameters as those in the $^{235}$U case. These parameters are fitted
to $\nup$, $\nud$, and cumulative FPY of $^{ 83}$Kr, $^{ 85}$Rb, $^{
86}$Kr, $^{ 87}$Sr, $^{131}$Xe, $^{132}$Xe, $^{133}$Xe, $^{134}$Xe,
$^{137}$Ba, $^{142}$Ce, $^{143}$Pr, $^{144}$Nd, $^{145}$Nd,
$^{146}$Nd, $^{147}$Nd, $^{148}$Nd, and $^{150}$Nd. They were chosen
from the mass chain evaluation by England and
Rider~\cite{LA-UR-94-3106}, where relatively small uncertainties are
assigned. The prior and posterior model parameters are given in
Table~\ref{tbl:parameter239}, and the comparison of mass yields are in
Fig.~\ref{fig:ya} (b). Similar to the $^{235}$U case, the prior
parameter set produces $\nud=0.00888$, which is too large compared to
the evaluated value of 0.00645. The smaller $\nud$ is achieved by the
adjustment, and the posterior set gives 0.00665.

\begin{table}
  \caption{Prior and posterior model parameters for $^{239}$Pu. See Table~\ref{tbl:parameter235} for parameter descriptions.}
  \label{tbl:parameter239}
  \begin{center}
    \begin{tabular}{crrrrrrrrrrrrrr}
\hline
           & pri   & post   & \multicolumn{11}{c}{uncertainty[\%] and correlation [\%]}  \\
\hline
$F_1$     &   0.234 &   0.248 &   4.1&$ 100$&      &      &      &      &      &      &      &      &      &      \\
$\sigma_1$&    3.51 &    3.26 &   5.2&$   2$&$ 100$&      &      &      &      &      &      &      &      &      \\
$\Delta_1$&    14.9 &    14.1 &   1.8&$   2$&$  18$&$ 100$&      &      &      &      &      &      &      &      \\
$F_2$     &   0.765 &   0.718 &   4.6&$  31$&$  -2$&$  -3$&$ 100$&      &      &      &      &      &      &      \\
$\sigma_2$&    6.06 &    6.58 &   0.5&$ -20$&$  21$&$  17$&$  20$&$ 100$&      &      &      &      &      &      \\
$\Delta_2$&    20.8 &    20.1 &   0.5&$  40$&$ -20$&$  -9$&$ -41$&$ -66$&$ 100$&      &      &      &      &      \\
$f_{Z0}$  &    1.00 &    2.58 &   4.4&$   1$&$   0$&$  69$&$  -1$&$ -21$&$  30$&$ 100$&      &      &      &      \\
$f_{N0}$  &    1.00 &    0.93 &  21.3&$  -1$&$   0$&$  -4$&$   1$&$   1$&$   0$&$   2$&$ 100$&      &      &      \\
$R_{T0}$  &    1.20 &    1.30 &   2.4&$  -6$&$  22$&$  10$&$   6$&$  29$&$ -69$&$ -31$&$  -2$&$ 100$&      &      \\
$f_J$     &    2.50 &    1.58 &   5.7&$ -24$&$  10$&$ -24$&$  24$&$  10$&$ -33$&$  14$&$   1$&$  28$&$ 100$&      \\
TKE       &   178.2 &   179.4 &   0.1&$  26$&$ -21$&$   1$&$ -26$&$  -7$&$  31$&$ -28$&$   0$&$ -26$&$ -92$&$ 100$\\
\hline
    \end{tabular}
  \end{center}
\end{table}


\subsubsection{Parameter adjustment for $^{238}$U}

Because fission observable data for $^{238}$U are only available in
the fast energy range and above, the procedure is slightly different
from the $^{235}$U and $^{239}$Pu cases. The adjusted Gaussian
parameters were obtained at 1.1 and 1.25~MeV by
Goverdovskiy~\cite{Goverdovskiy2000} and 1.2~MeV~\cite{Vives2000} by
Vives. The adjusted Gaussian parameters are
\begin{eqnarray}
   \Delta_1 &=& -\Delta_5 =  15.66 - 0.0679 E_n \ ,\\
   \Delta_2 &=& -\Delta_4 =  23.36 - 0.1929 E_n \ ,\\
   \sigma_1 &=& \sigma_5  =   3.31 + 0.0159 E_n \ ,\\
   \sigma_2 &=& \sigma_4  =   5.43 + 0.1267 E_n \ ,\\
   \sigma_3 &=& 4.50 + 0.267100 E_n \ .
\end{eqnarray}
The fractions of each Gaussian are given by
\begin{eqnarray}
   F_1 &=& F_5 =  0.5876 + 0.032 E_n \ , \\
   F_2 &=& F_4 =  0.4203 - 0.034 E_n \ ,\\
   F_3 &=&        0.0006 + 0.001 E_n \ ,
   \label{eq:FjU238}
\end{eqnarray}
and the covariance matrix is given in Table~\ref{tbl:parameter238}.
These parameters are fitted to $\nup$, $\nud$, and cumulative FPY of
$^{ 97}$Zr, $^{133}$I, $^{135}$Xe, $^{137}$Cs, $^{140}$Ba, $^{143}$Ce,
$^{14}$Ce, $^{145}$Pr,$^{147}$Nd, and $^{148}$Nd.

\begin{table}
  \caption{Prior and posterior model parameters for $^{238}$U at $\approx$1~MeV. See Table~\ref{tbl:parameter235} for parameter descriptions.}
  \label{tbl:parameter238}
  \begin{center}
    \begin{tabular}{crrrrrrrrrrrrrr}
\hline
           & pri    & post & \multicolumn{11}{c}{uncertainty[\%] and correlation [\%]}  \\
\hline
$F_1$      &   0.587 &   0.625 &   3.6&$ 100$&      &      &      &      &      &      &      &      &      &      \\
$\sigma_1$ &   5.405 &   5.580 &   1.4&$  16$&$ 100$&      &      &      &      &      &      &      &      &      \\
$\Delta_1$ &  22.879 &  23.128 &   0.5&$ -43$&$ -18$&$ 100$&      &      &      &      &      &      &      &      \\
$F_2$      &   0.413 &   0.380 &   4.4&$  33$&$ -10$&$  55$&$ 100$&      &      &      &      &      &      &      \\
$\sigma_2$ &   3.459 &   3.326 &   2.6&$ -40$&$   8$&$  42$&$  31$&$ 100$&      &      &      &      &      &      \\
$\Delta_2$ &  15.515 &  15.584 &   0.7&$ -11$&$  27$&$  43$&$  30$&$  13$&$ 100$&      &      &      &      &      \\
$f_{Z0}$   &   1.00  &   2.386 &   5.3&$   0$&$ -16$&$  25$&$  20$&$ -14$&$  48$&$ 100$&      &      &      &      \\
$f_{N0}$   &   1.00  &   0.736 &  52.8&$   0$&$  -8$&$  -2$&$   0$&$  -2$&$   3$&$  13$&$ 100$&      &      &      \\
$R_{T0}$   &   1.30  &   1.327 &   1.0&$  -1$&$  -1$&$  -6$&$   2$&$   0$&$  -9$&$  22$&$   0$&$ 100$&      &      \\
$f_J$      &   3.00  &   2.956 &   1.0&$   3$&$   6$&$  -6$&$  -3$&$   1$&$  -4$&$  -4$&$   0$&$   2$&$ 100$&      \\
TKE        & 171.4  &   170.5 &   0.1&$  -5$&$ -29$&$  -9$&$   5$&$  -2$&$ -30$&$  28$&$   0$&$  -4$&$ -18$&$ 100$\\
\hline     
    \end{tabular}
  \end{center}
\end{table}

\subsection{Energy dependence of $\nup$ and $\nud$}
\subsubsection{Energy-dependent inputs and pivots}

Some of the Gaussian parameters are weakly energy-dependent, and often
expressed by a linear function of the incident energy as in
Eqs.~(\ref{eq:FjPu239a}) -- (\ref{eq:FjPu239c}).  The energy-dependent terms are obtained by
fitting to the experimental $Y_P(A)$ data, and we do not attempt to
tune these parameters. We consider other parameters, $R_T$, $f_Z$,
$f_N$, and TKE, to be energy-dependent, and simple linear functions
are assumed as in Eqs.(\ref{eq:RT}), (\ref{eq:FZ}) and
(\ref{eq:FN}). Since the energy-dependence of TKE is rather well know
experimentally, we study the energy-dependence of the FPYs, $\nup$,
and $\nud$ by assuming a simple form for the mode inputs for $^{235}$U
and $^{239}$Pu first. We exclude $^{238}$U for now, as it is a
threshold fissioner.  The $R_{T1}$ parameter in Eq.~(\ref{eq:RT}) is
roughly $ - (R_T(0)-1) / 6.0$ MeV$^{-1}$ to make $R_T=1$ at the
opening of second chance fission, hence $R_{T1} = -0.0476$ and
$-0.0507$ for $^{235}$U and $^{239}$Pu. Similarly, $f_{Z1}$ and
$f_{N1}$ are estimated to be $f_{Z1}=-0.296$~MeV$^{-1}$ and
$f_{N1}=-0.161$~MeV$^{-1}$ for $^{235}$U, and
$f_{Z1}=-0.430$~MeV$^{-1}$ and $f_{N1}=-0.156$~MeV$^{-1}$ for
$^{239}$Pu, which ensures that the even-odd effect disappears at
$E_n=6$~MeV.

First we consider four cases; (1) both $R_T$ and $f_{Z,N}$ are
constant, (2) constant $R_T$ and energy-dependent $f_{Z,N}$, (3)
energy-dependent $R_T$ and constant $f_{Z,N}$, and (4) both
energy-dependent. By comparing the calculated $\nup$ and $\nud$ with
experimental data, we found that the energy-dependence of $R_T$
modestly impacts on the results, and probably the modeling uncertainty
conceals the importance of $R_T$. Whereas we also noticed that the
energy-dependence of $f_{Z,N}$ is crucial for $\nud$. Hereafter we
assume $R_T$ is constant, while $f_{Z,N}$ is energy-dependent.

\begin{figure}
  \begin{center}
    \resizebox{0.8\textwidth}{!}{\includegraphics{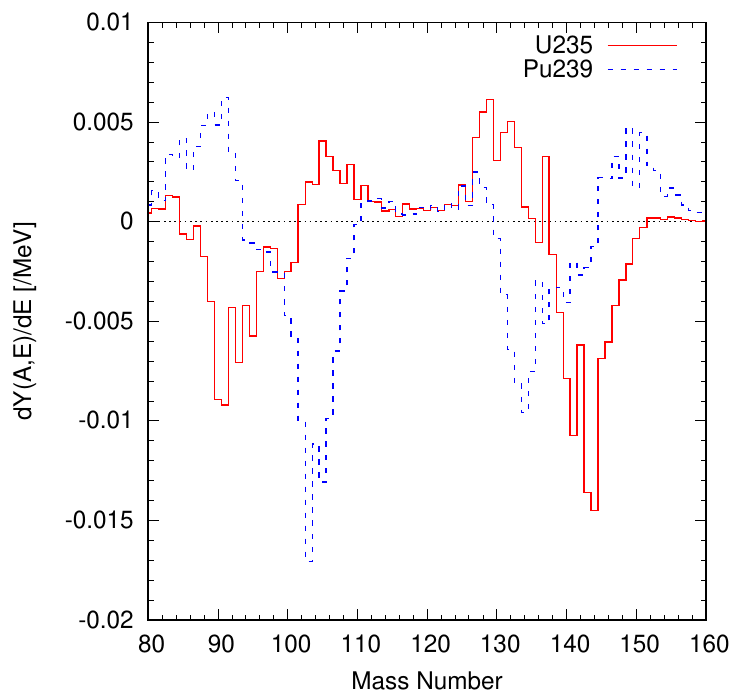}} 
  \end{center}
  \caption{Energy-dependence of the cumulative mass yields,
    $\partial Y(A,E)/\partial E$, approximated by $(Y(A,2\mbox{[MeV]})-Y(A,0\mbox{[MeV]}))/2$.}
  \label{fig:massyield-diff}
\end{figure}

When an independent or cumulative FPY is almost energy-independent,
\begin{equation}
 \frac{d Y_{I,C}(Z,A,E)}{dE} \simeq 0 \ ,
\end{equation}
it is easier to see the mass region where this condition happens by 
calculating the derivative of mass yields,
\begin{equation}
 \frac{d Y_{I,C}(A,E)}{dE} = \sum_Z \frac{d Y_{I,C}(Z,A,E)}{dE} \simeq 0 \ .
 \label{eq:dYdE}
\end{equation}
We approximate the derivative by coarse numerical derivative
$(Y_{I,C}(A,2 \mbox{[MeV]}) - Y_{I,C}(A,0 \mbox{[MeV]}))/2$, which is
shown in Fig.~\ref{fig:massyield-diff}.  The general shape of
$dY_I/dE$ does not change too much in the energy range below the
second chance fission. This implies the cumulative FPYs vary
monotonously with the incident neutron energy.

The derivative plot for $^{235}$U indicates FPYs near $A=85$, 100, and
135 vary slowly with the energy, while FPYs near $A=90$, 104, 129, and
143 should have steeper energy-dependence. These energy-independent
regions, or the pivots, appear due to complicated interplay among the
energy-dependent model parameters. In the case of $^{239}$Pu, the
pivots locate near $A=92$, 109, 129, and 142, and the FPYs in the peak
regions ($A=103$ and 133) may show the largest reduction rate.

\begin{figure}[htbp]
 \begin{minipage}{0.33\hsize}
  \begin{center}
    $^{235}$U\\
    \resizebox{1.0\textwidth}{!}{\includegraphics{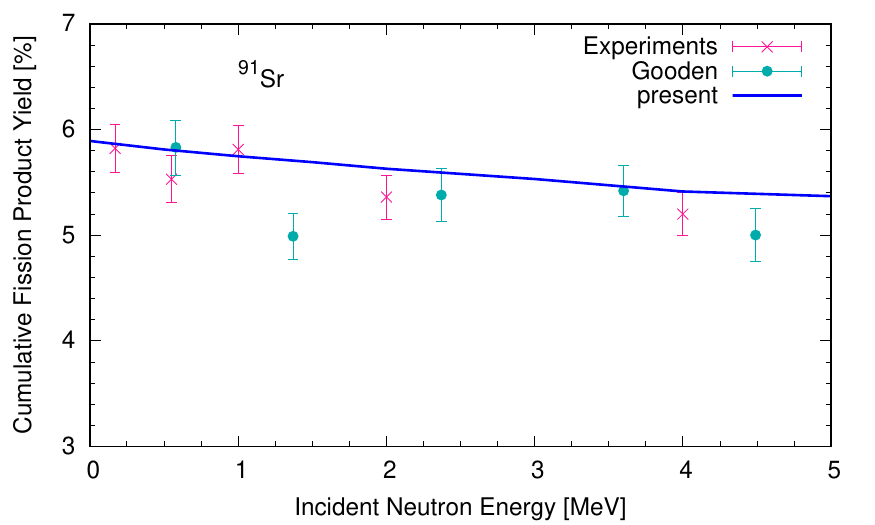}}\\
    \resizebox{1.0\textwidth}{!}{\includegraphics{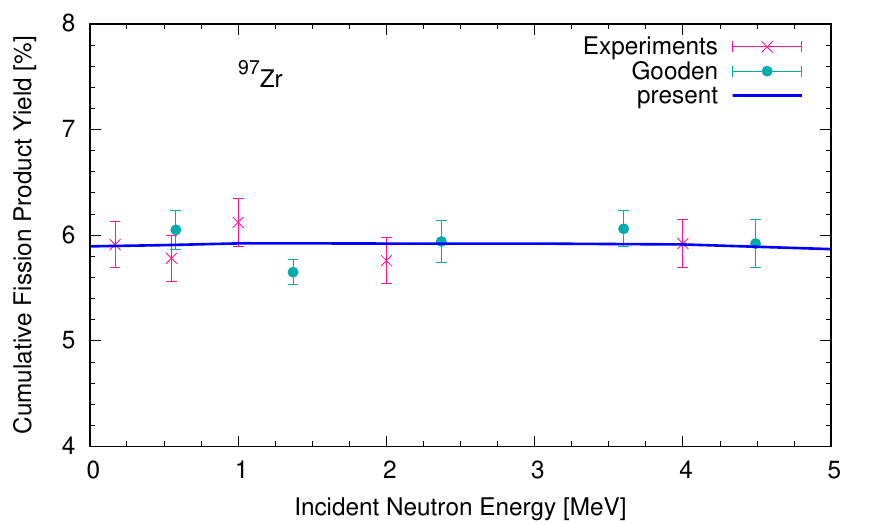}}\\
    \resizebox{1.0\textwidth}{!}{\includegraphics{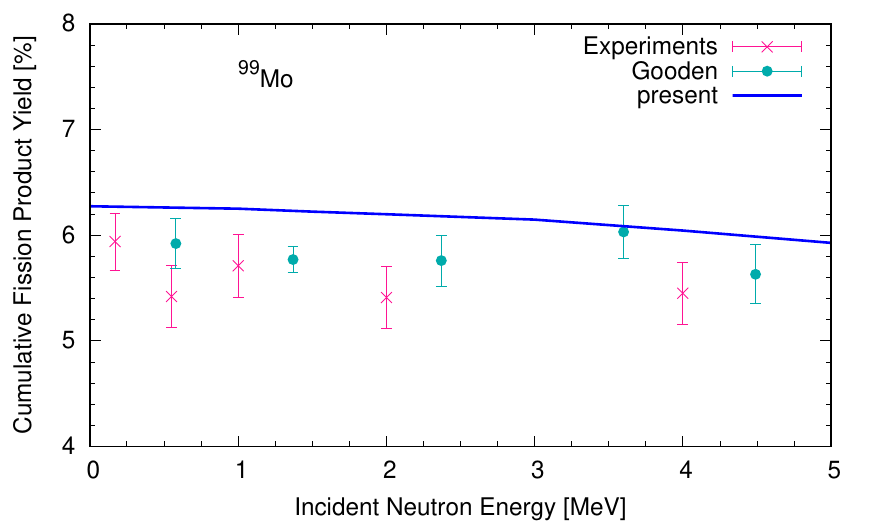}}\\
    \resizebox{1.0\textwidth}{!}{\includegraphics{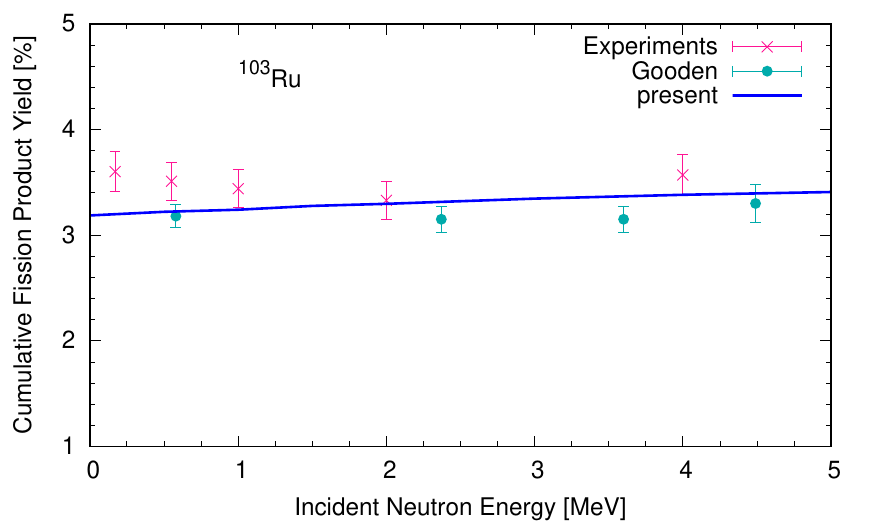}}\\
  \end{center}
\end{minipage}
\begin{minipage}{0.33\hsize}
  \begin{center}
    $^{239}$Pu\\
    \resizebox{1.0\textwidth}{!}{\includegraphics{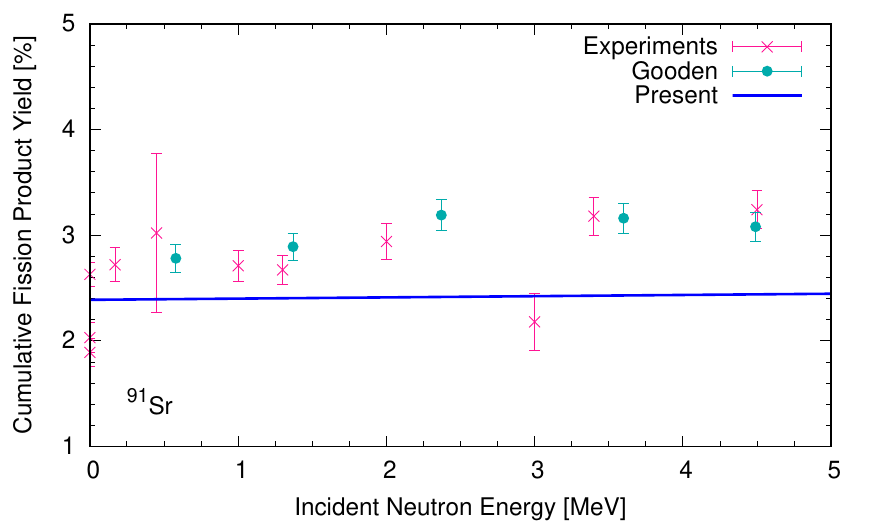}}\\
    \resizebox{1.0\textwidth}{!}{\includegraphics{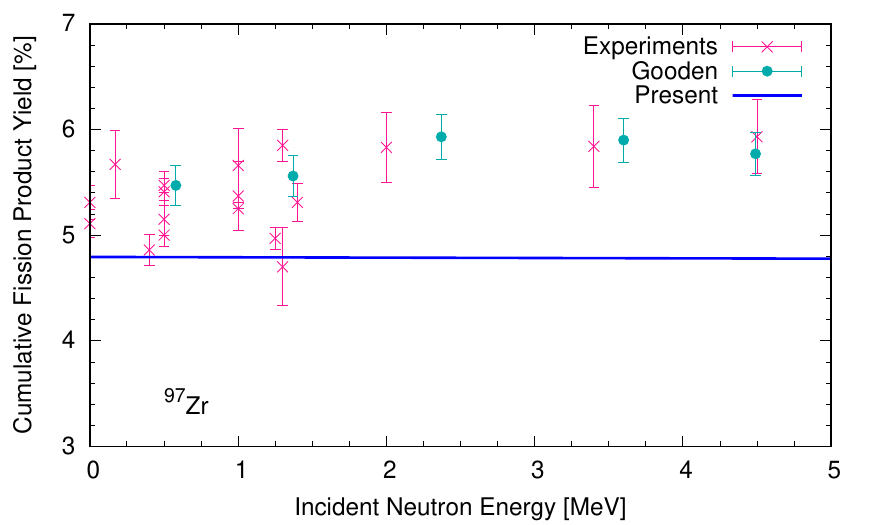}}\\
    \resizebox{1.0\textwidth}{!}{\includegraphics{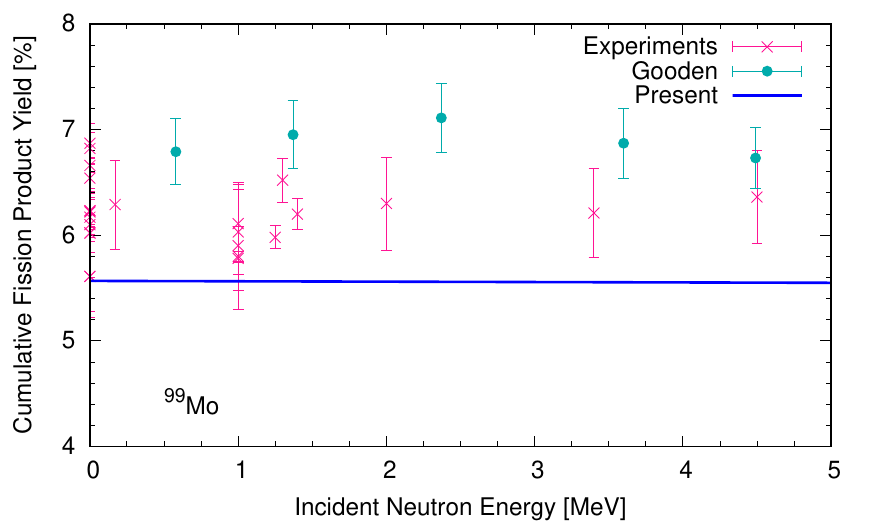}}\\
    \resizebox{1.0\textwidth}{!}{\includegraphics{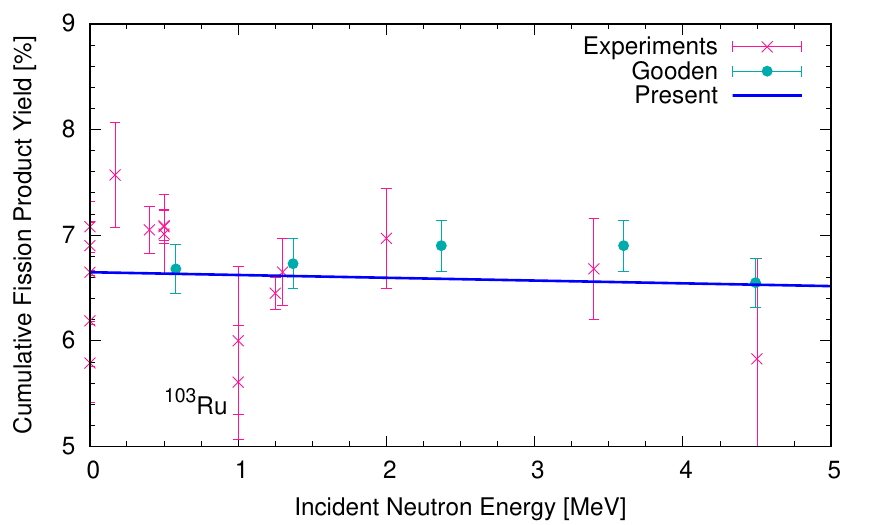}}\\
  \end{center}
\end{minipage}
\begin{minipage}{0.33\hsize}
  \begin{center}
    $^{238}$U\\
    \resizebox{1.0\textwidth}{!}{\includegraphics{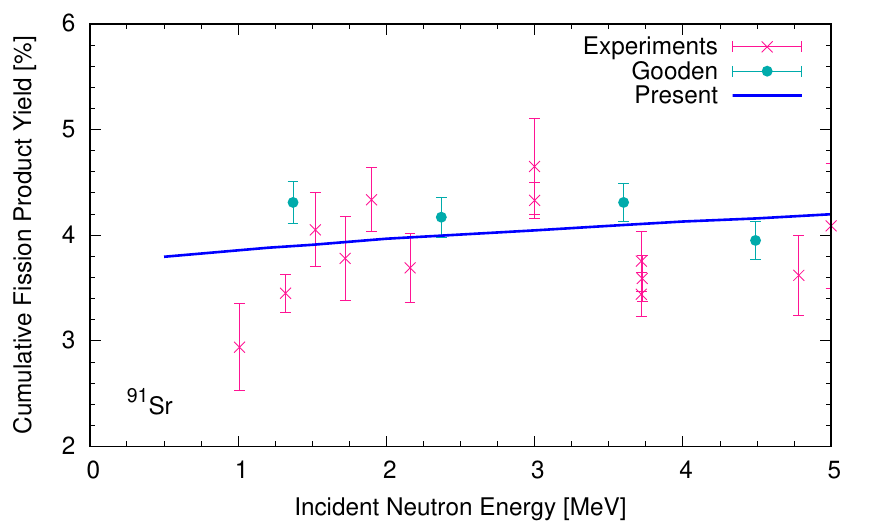}}\\
    \resizebox{1.0\textwidth}{!}{\includegraphics{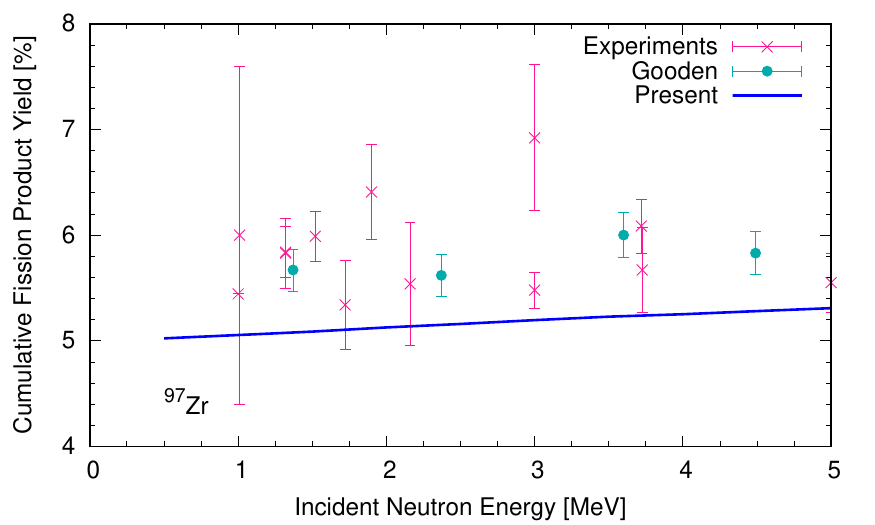}}\\
    \resizebox{1.0\textwidth}{!}{\includegraphics{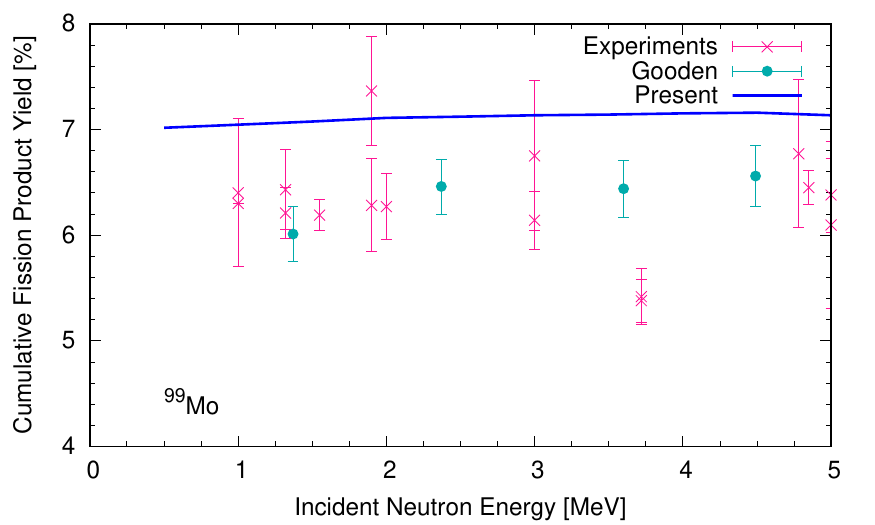}}\\
    \resizebox{1.0\textwidth}{!}{\includegraphics{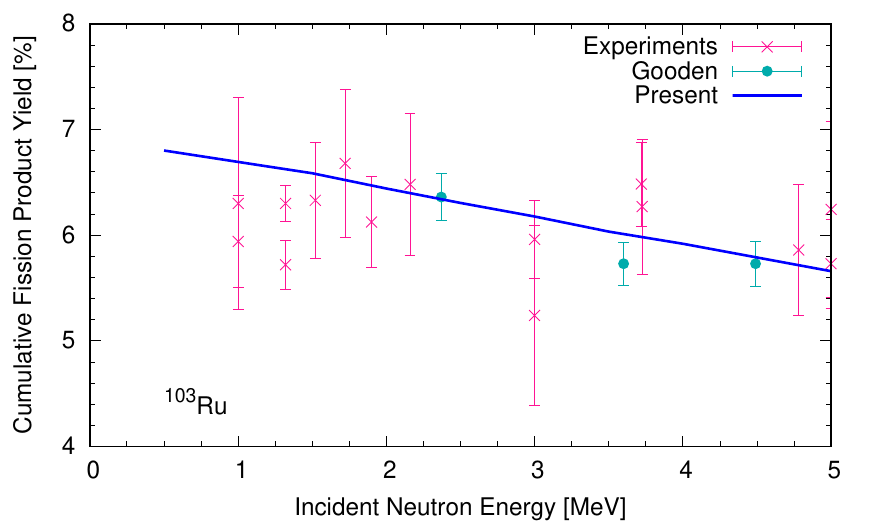}}\\
  \end{center}
\end{minipage}
  
  \caption{Energy dependence of cumulative fission product yields of
           $^{91}$Sr, $^{97}$Zr, $^{99}$Mo, and $^{103}$Ru for the
           neutron-induced fission on $^{235}$U (left), $^{239}$Pu
           (middle), and $^{238}$U (right) of calculated data (solid
           line) compared with with the experimental data of Gooden et
           al.~\cite{Gooden2016}, as well as other published data.}
\label{fig:U235cfpy1}
\end{figure}

In Fig.~\ref{fig:U235cfpy1} we compare some of our calculated
$Y_C(Z,A,E)$ with the experimental data of Gooden et
al.~\cite{Gooden2016}, measurements at LANL in the critical
assemblies~\cite{Chadwick2010}, as well as other published data.  From
the derivative plot in Fig.~\ref{fig:massyield-diff}, we expect $Y_C$
of $^{235}$U decreases in the $A\simeq 90$ region, while $Y_C$ of
$^{239}$Pu increases. For the both isotopes, the pivot will be seen in
$A=95$ -- 100. The comparisons of $^{91}$Sr, $^{97}$Zr, and $^{99}$Mo
clearly show these behavior, and $^{103}$Ru now shows an opposite
tendency as the incident neutron energy.

\begin{figure}
 \begin{minipage}{0.33\hsize}
  \begin{center}
    $^{235}$U\\
    \resizebox{1.0\textwidth}{!}{\includegraphics{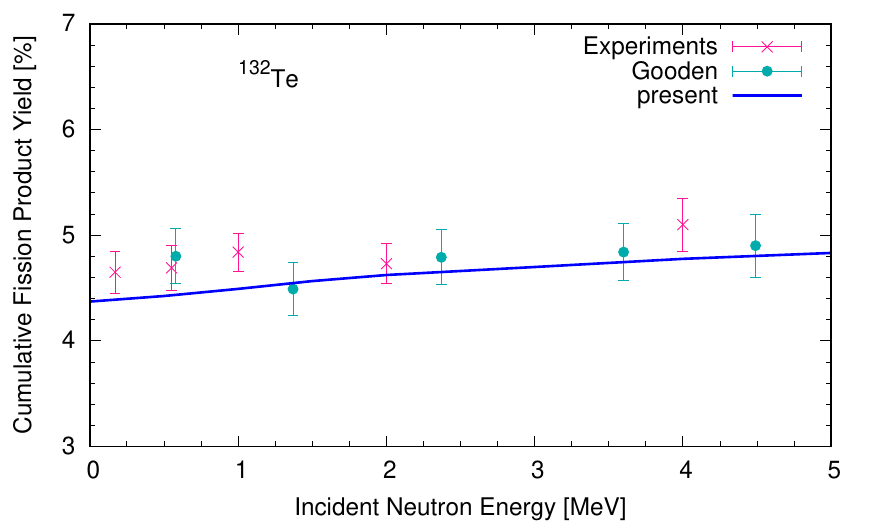}}\\
    \resizebox{1.0\textwidth}{!}{\includegraphics{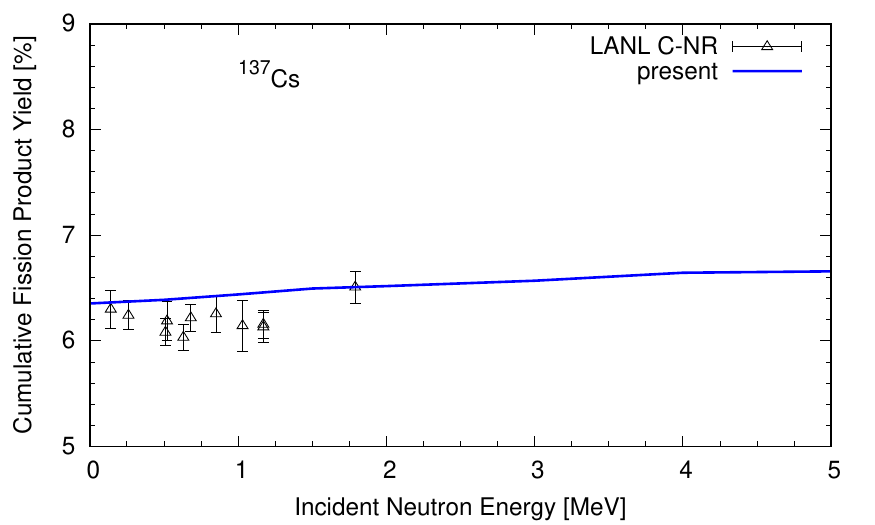}}\\
    \resizebox{1.0\textwidth}{!}{\includegraphics{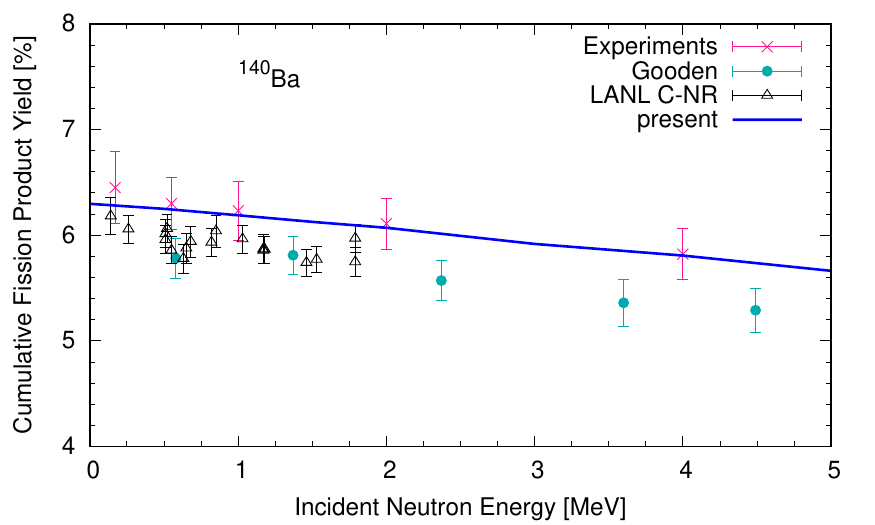}}\\
    \resizebox{1.0\textwidth}{!}{\includegraphics{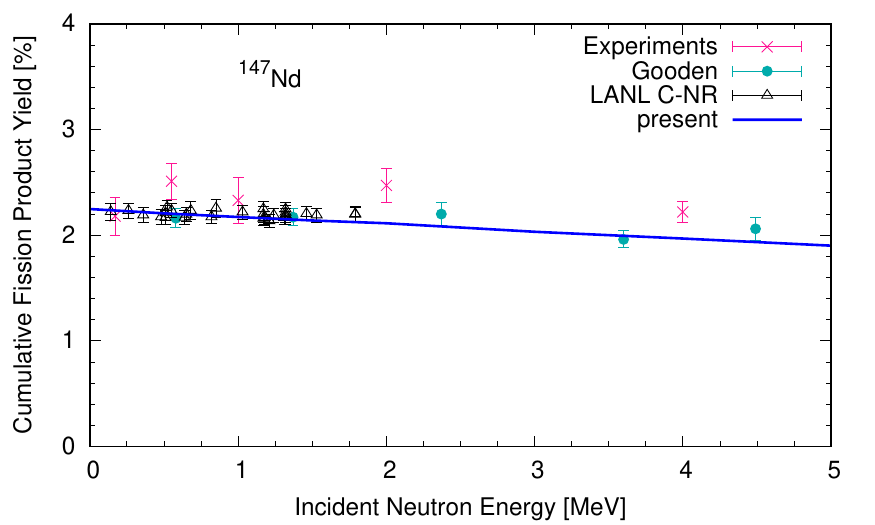}}\\
  \end{center}
\end{minipage}
\begin{minipage}{0.33\hsize}
  \begin{center}
    $^{239}$Pu\\
    \resizebox{1.0\textwidth}{!}{\includegraphics{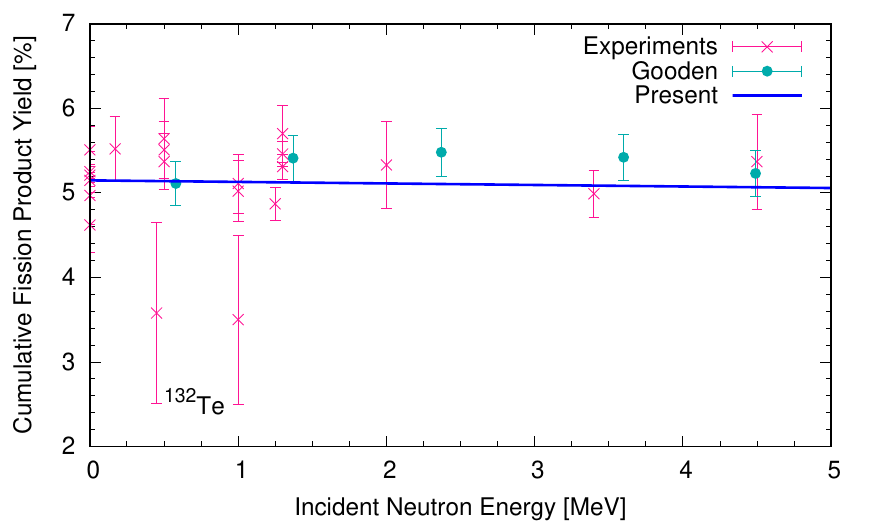}}\\
    \resizebox{1.0\textwidth}{!}{\includegraphics{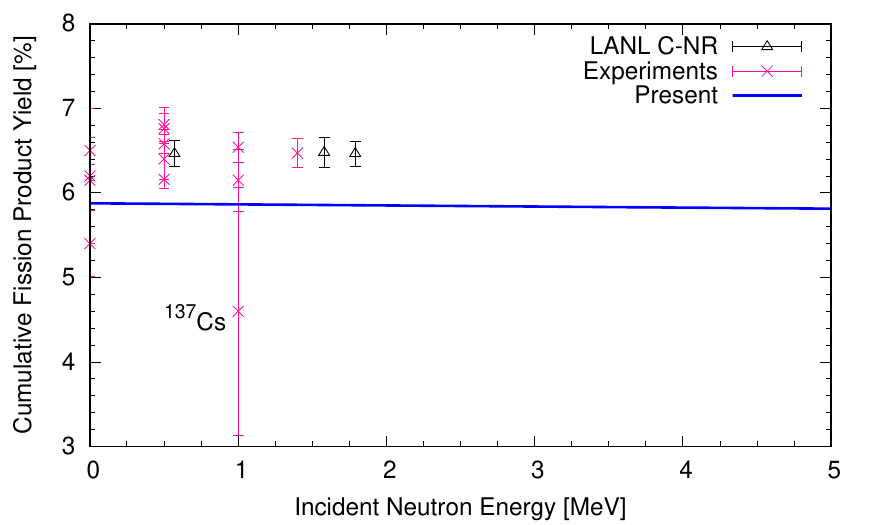}}\\
    \resizebox{1.0\textwidth}{!}{\includegraphics{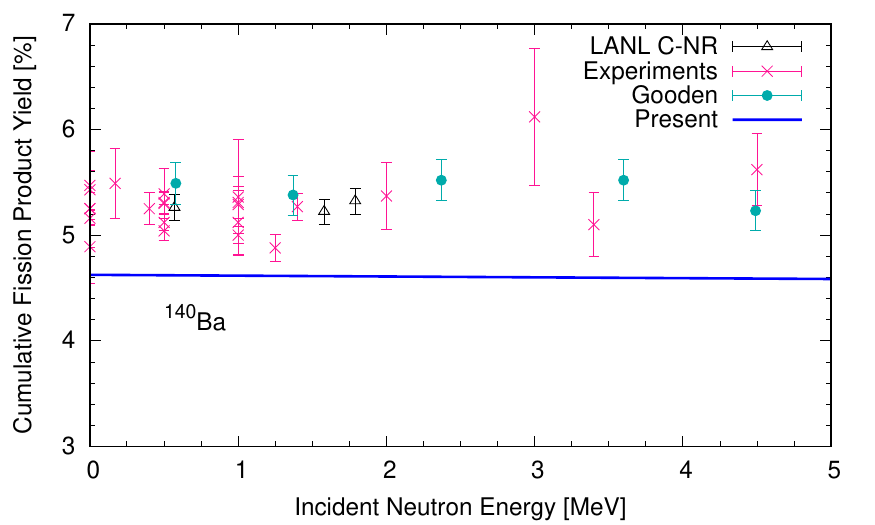}}\\
    \resizebox{1.0\textwidth}{!}{\includegraphics{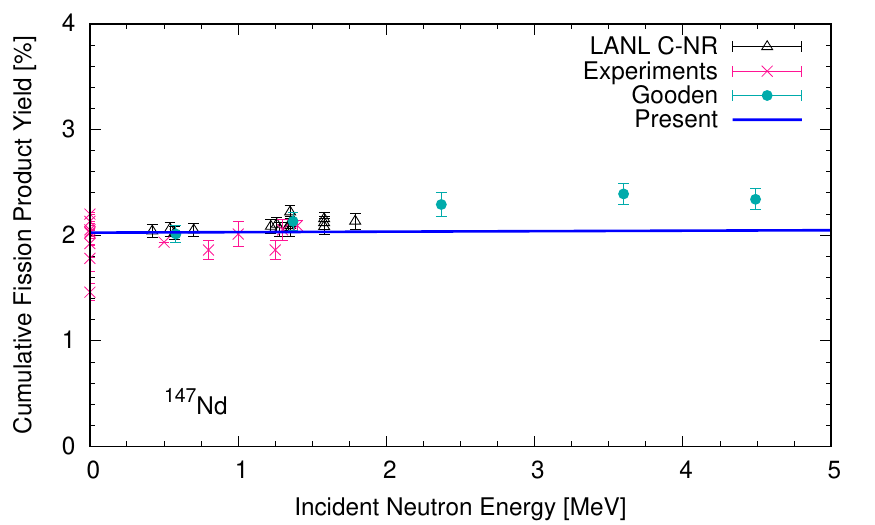}}\\
  \end{center}
\end{minipage}
\begin{minipage}{0.33\hsize}
  \begin{center}
    $^{238}$U\\
    \resizebox{1.0\textwidth}{!}{\includegraphics{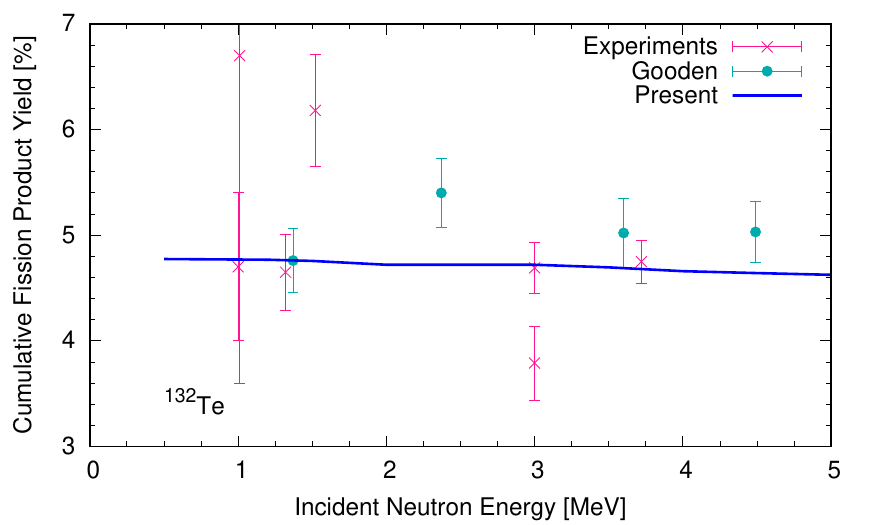}}\\
    \resizebox{1.0\textwidth}{!}{\includegraphics{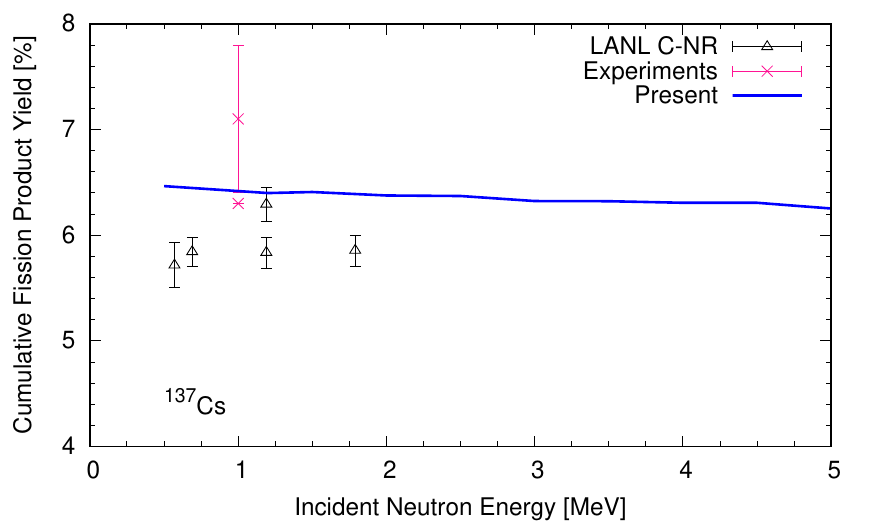}}\\
    \resizebox{1.0\textwidth}{!}{\includegraphics{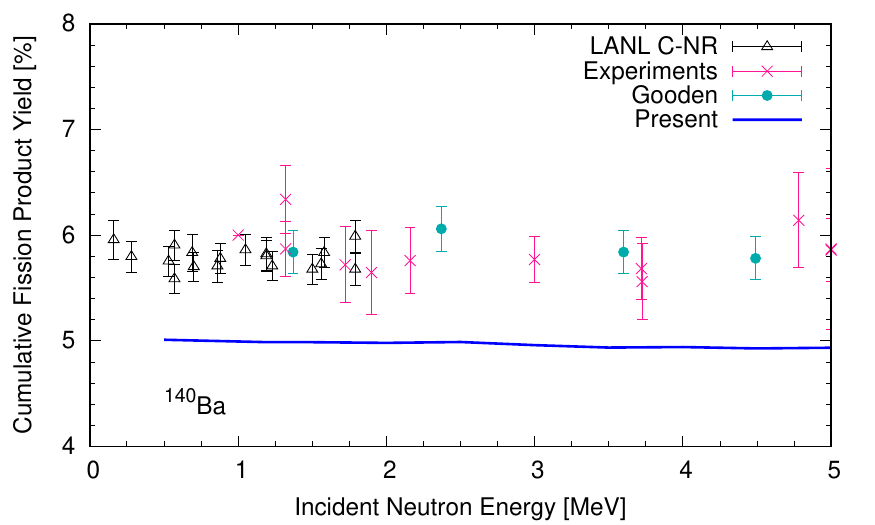}}\\
    \resizebox{1.0\textwidth}{!}{\includegraphics{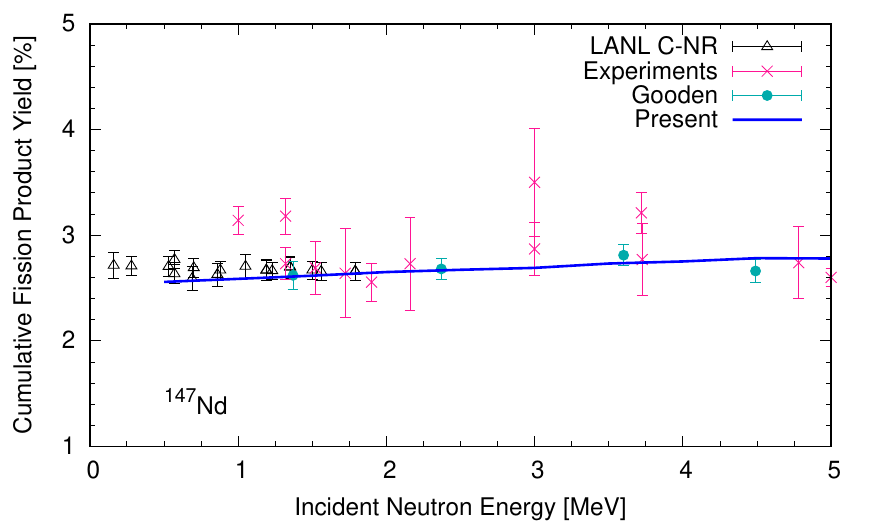}}\\
  \end{center}
\end{minipage}
  \caption{Incident neutron energy dependence of cumulative fission product yields of
            $^{132}$Te, $^{137}$Cs, $^{140}$Ba, and $^{147}$Nd for the neutron-induced
           fission on $^{235}$U (left), $^{239}$Pu
           (middle), and $^{238}$U (right) of calculated data (solid
           line) compared with with the experimental data of Gooden et
           al.~\cite{Gooden2016} and measurements at LANL in the
           critical assemblies (LANL CN-R)~\cite{Chadwick2010}, as
           well as other published data.}
  \label{fig:U235cfpy2}
\end{figure}

On the heavier mass side, the slope of
$Y_C(Z,A,E)$ changes the sign from positive to negative around
$A=134$ for $^{235}$U, with one exception of the $A=137$ case that has the positive
slope. For $^{239}$Pu, the sign change happens twice, near $A=130$ and 145.
This is shown in Fig.~\ref{fig:U235cfpy2}; $^{132}$Te,
$^{137}$Cs, $^{140}$Ba, and $^{147}$Nd. Our model calculation also
reproduces other isotopes with the similar quality.

Although we didn't include the $^{238}$U case in Fig.~\ref{fig:massyield-diff} 
as FPY at thermal is only given by extrapolation,
Figs.~\ref{fig:U235cfpy1} and \ref{fig:U235cfpy2} include $^{238}$U too.

\subsubsection{Energy-dependence of $\nup$ and $\nud$}

\begin{figure}
  \begin{center}
    \resizebox{0.48\textwidth}{!}{\includegraphics{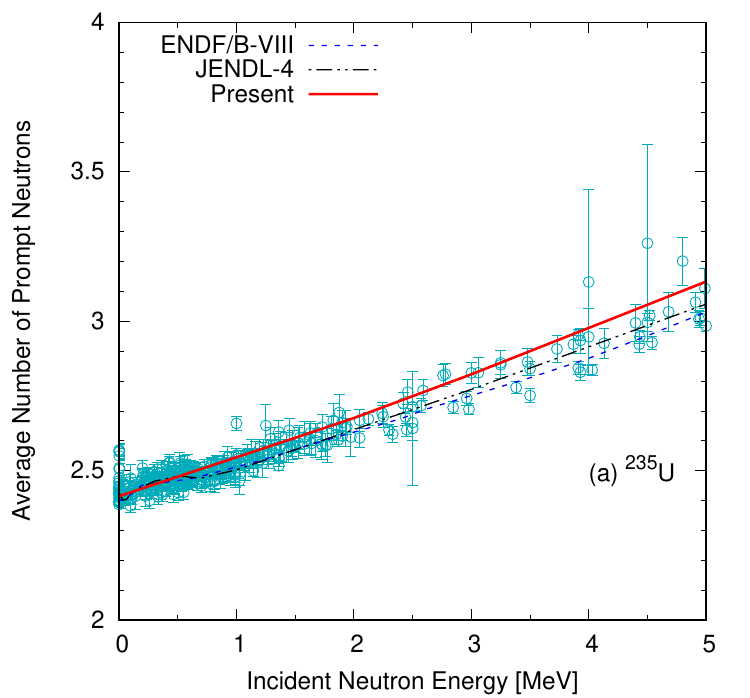}}\\
    \resizebox{0.48\textwidth}{!}{\includegraphics{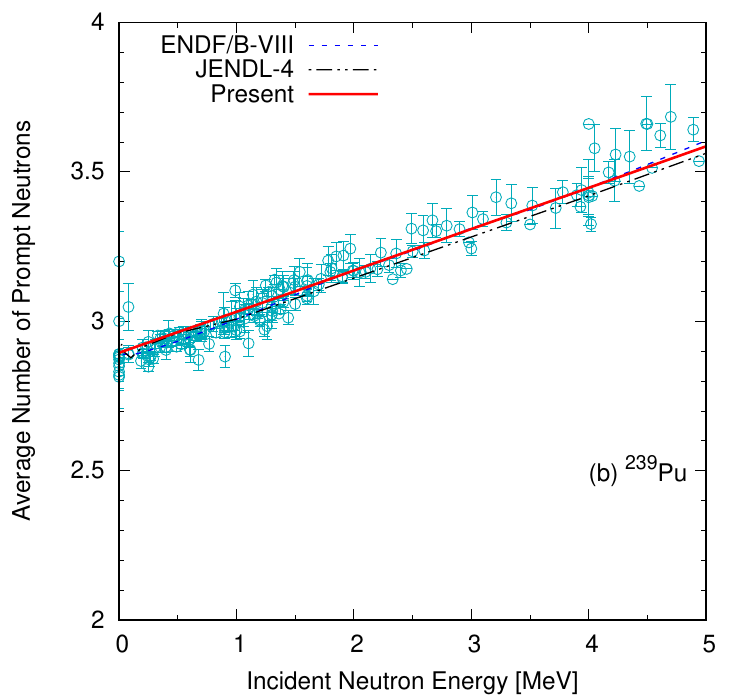}}\\
    \resizebox{0.48\textwidth}{!}{\includegraphics{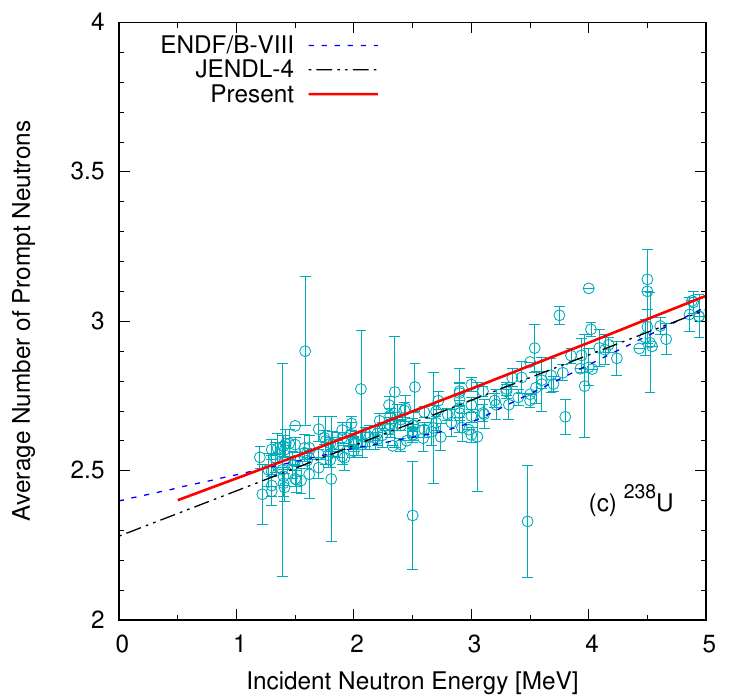}}
  \end{center}
  
  \caption{Calculated incident neutron energy dependence of $\nup$ for
  $^{235}$U,$^{239}$Pu, and $^{238}$U (solid line) compared with the
  evaluated $\nup$ in ENDF/B-VIII (dotted line) and JENDL-4.0
  (dot-dashed line) and available experimental data.}
  
  \label{fig:nup}
\end{figure}

\begin{figure}
  \begin{center}
    \resizebox{0.48\textwidth}{!}{\includegraphics{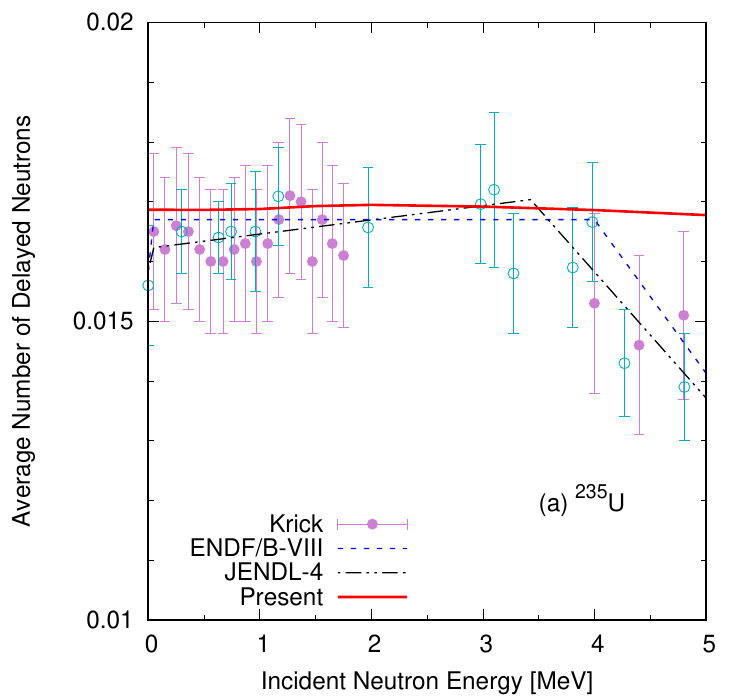}}\\
    \resizebox{0.48\textwidth}{!}{\includegraphics{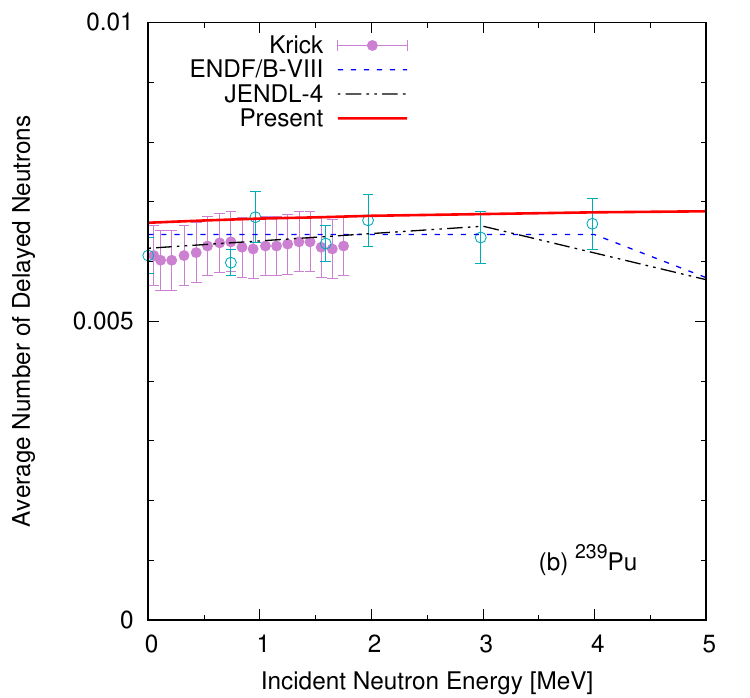}}\\
    \resizebox{0.48\textwidth}{!}{\includegraphics{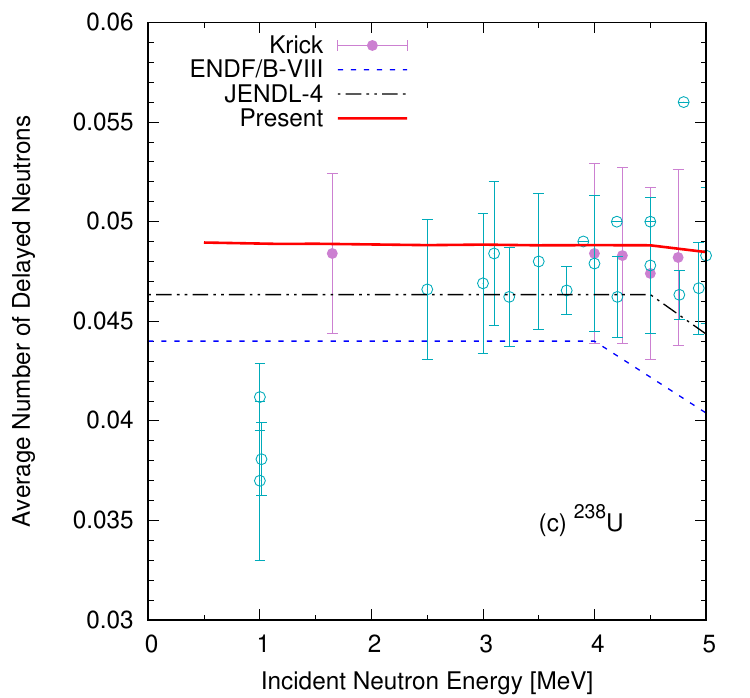}}
  \end{center}
  
  \caption{Calculated incident neutron energy dependence of $\nud$ for
  $^{235}$U,$^{239}$Pu, and $^{238}$U (solid line) compared with the
  evaluated $\nud$ in ENDF/B-VIII (dotted line) and JENDL-4.0
  (dot-dashed line), and experimental measurements by Krick {\it
  et.al}~\cite{Krick1972} (filled circle) and the other available
  experimental data (open circle).}
  
  \label{fig:nud}
\end{figure}

The calculated $\nup$ and $\nud$ for $^{235,238}$U and $^{239}$Pu are
compared with experimental data in Figs.~\ref{fig:nup} and
\ref{fig:nud}. The evaluated $\nup$ and $\nud$ in ENDF/B-VIII and
JENDL-4.0, which are evaluated by least-squares fitting to the available
experimental data, are also compared.  In general $\nup$ increases as
the incident neutron energy goes higher, simply because of the energy
conservation. However, its slope $d\nup/dE$ strongly depends on the
behavior of TKE. Although the mechanism for the incident-energy
dependence of TKE is still unclear, we take the energy dependence of
TKE from experimental data, and it enables us to reproduce $\nup$ by
our model.  Other parameters, the Gaussian shape and $f_{Z,N}$, also
change the slope of $\nup$, but they have a much more modest impact on the
calculated result.

The energy-dependence of $\nud$ is caused mainly by changing the yields of the
delayed neutron precursors. Interestingly the calculated and
experimental $\nud$'s reveal very weak energy-dependency for these
isotopes. As we noted large fractions of delayed neutron emission are
from the mass regions of $A=137$ and 94, and according to
Fig.~\ref{fig:massyield-diff}, we expect $\nud$ to decrease.

\begin{figure}
  \begin{center}
    \resizebox{0.48\textwidth}{!}{\includegraphics{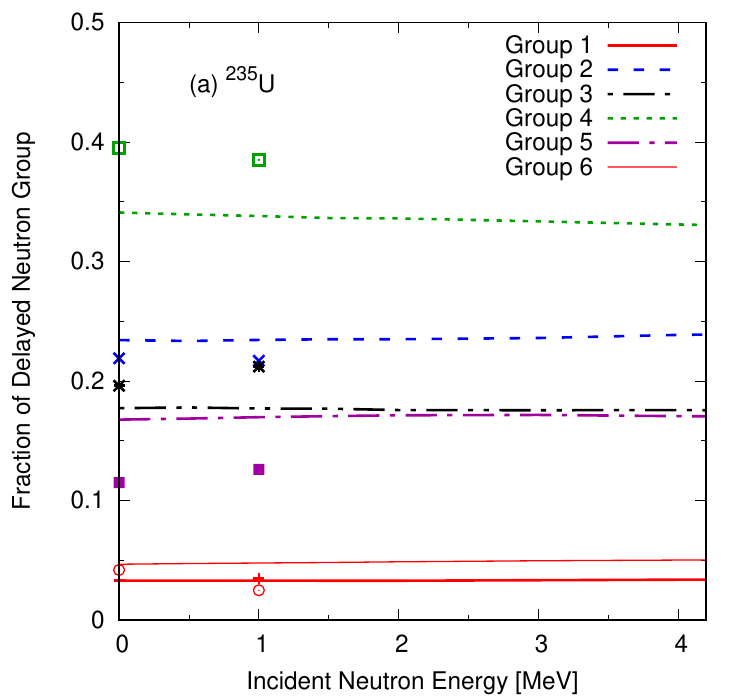}}\\
    \resizebox{0.48\textwidth}{!}{\includegraphics{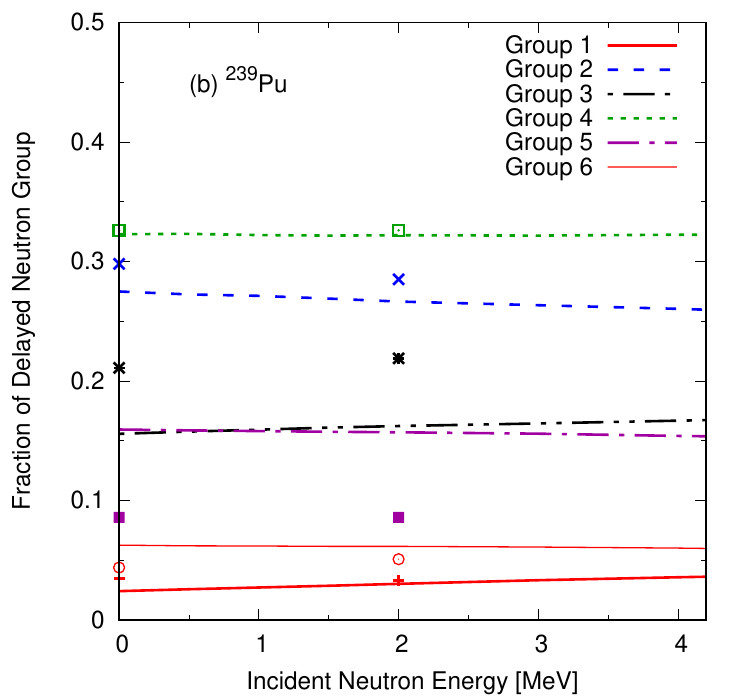}}\\
    \resizebox{0.48\textwidth}{!}{\includegraphics{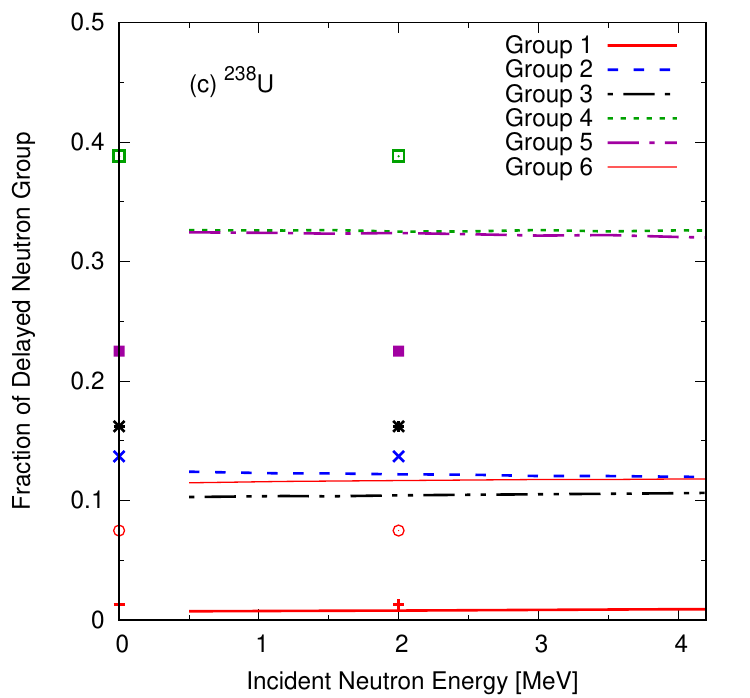}}
  \end{center}

  \caption{Calculated relative contribution to $\nud$ from each of the
           delayed neutron groups for $^{235}$U,$^{239}$Pu, and
           $^{238}$U.  Evaluated $\nud$ fractions were taken from
           JENDL-4.0 and are shown by the symbols ($+$:Group 1,
           $\times$: Group 2, $\ast$: Group 3, $\square$: Group 4,
           $\blacksquare$: Group 5, and $\bigcirc$: Group 6) at
           thermal and 1 MeV for $^{235}$U, and at thermal and 2 MeV
           for $^{239}$Pu, and $^{238}$U.}

  \label{fig:nudgroup}
\end{figure}

As it is not so convenient to survey the delayed neutron precursors
individually, we lump the precursors into the well-known six groups
according to their half-lives $T_{1/2}$, and calculate the
energy-dependence of the six-group yields. The group structure is
usually defined by the isotopes included in each group. This is
convenient for the longer $T_{1/2}$ groups, but it is ambiguous for
the shorter groups.  For the sake of convenience, we define the
six-group structure as (1) $T_{1/2} > 40$~s, (2) $8 < T_{1/2} \le
40$~s, (3) $3 < T_{1/2} \le 8$~s, (4) $1 < T_{1/2} \le 3$~s, (5) $0.3
< T_{1/2} \le 1$~s, and (6) $T_{1/2} \le 0.3$~s. The fractions of each
group are shown in Fig.~\ref{fig:nudgroup}. In the case of $^{235}$U,
the largest contribution is from the Group 4, which slightly decreases
as the incident neutron energy.  This is compensated by the increasing
Group 2, resulting in the flat behavior of $\nud$. The energy
variation of each group is more visible for the $^{239}$Pu
case. Obviously the energy-dependence of $\nud$ does not originate
from specific fission products, but a consequence of their
competition.

We studied sensitivities of the model parameters to $d\nud/dE$, and
found that the $f_{Z1}$ and $f_{N1}$ terms change the slope.  When
$f_{Z1}=f_{N1}=0$, or a constant odd-even effect, $\nud$ decreases for
both $^{235}$U and $^{239}$Pu cases. We briefly estimated
the energy-dependence of the odd-even term so that this effect fades
away toward the second chance fission. Nonetheless, this ansatz was not so
unrealistic. Better reproduction of the experimental data can be
achieved by adjusting the $f_{Z1}$ and $f_{N1}$ parameters, yet the
currently available data have rather large uncertainties to 
estimate these parameters precisely.

\subsection{Extrapolating to the second chance fission}
The experimental data of $\nud$ for $^{235}$U drop sharply near
5~MeV~\cite{Krick1972,Evans1973}, and the evaluated data often include
a curious kink to reproduce this behavior. As we demonstrated that
$\nud$ is weakly energy-dependent up to the second chance fission, the
kink could be hypothetically the evidence of the second-chance
contribution, namely transition of major fissioning system from
$^{236}$U to $^{235}$U. The full-extension of our FPY model by
including the multi-chance fission is underway~\cite{Lovell2021}, and
here we extrapolate our calculations beyond the second-chance fission
threshold. We do not intend to perform a detailed model parameter
adjustment as done for the first chance case, but similar parameters
obtained by the first-chance calculation were plugged into the
second-chance fission to see if we will be able to reproduce the
kink. This exercise is done for the $^{235}$U case only.

The fission probabilities $P_f(E)$ for the first and second chances
are calculated with the CoH$_3$ code~\cite{Kawano2019}. The fission
parameters, such as the fission barrier, curvature, and level density,
are adjusted to reproduce the evaluated fission cross section of
$^{235}$U. We use the same $Y_P(A)$ for the second chance, but
shifted the mid-point by 1/2 mass unit to the lower mass side. $f_J$,
TKE, and $f_{Z,N}$ for both $^{236}$U and $^{235}$U are the same.

\begin{figure}
  \begin{center}
    \resizebox{0.7\textwidth}{!}{\includegraphics{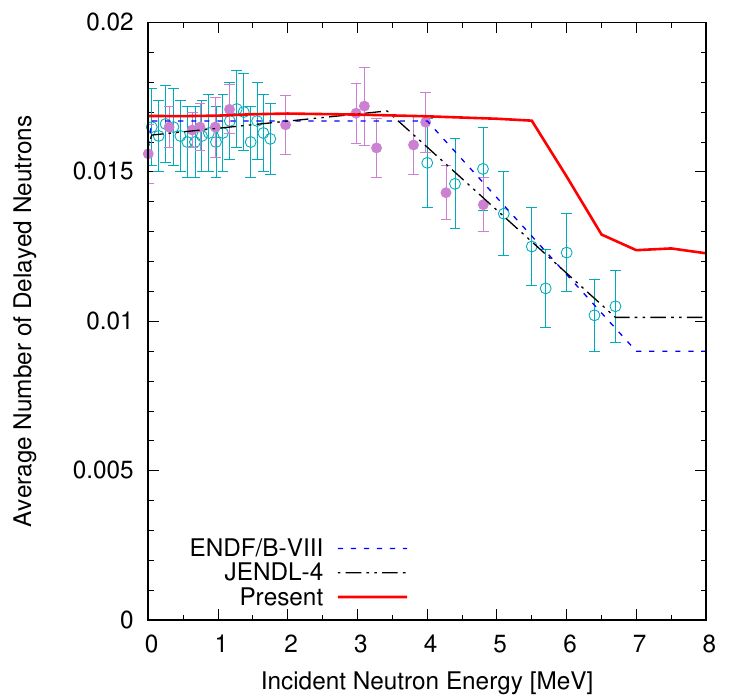}}
  \end{center}
  \caption{Energy dependence of $\nud$ for $^{235}$U when the second-chance
           fission is involved.}
  \label{fig:nudextend}
\end{figure}

The calculated $\nud$ is shown in Fig.~\ref{fig:nudextend}. Albeit the
calculated $\nud$ drops at the energy that is about 1.5~MeV higher
than the experimental data, the shape is well reproduced. This
supports our hypothesis of the transition of fissioning systems from
the first compound nucleus to the second one. At 8~MeV the
probability of second chance fission reaches 80\%, and a new set of 
delayed neutron emitters again forms a new plateau above that energy.
The step-function-like behavior of $\nud$ is thus understood.

The calculated transition energy, which is basically the second-chance fission 
threshold, is higher than the experimental data, and this is still an open question.
Despite the fact that our fission barrier parameters could
have some uncertainties, the 1.5-MeV change in the fission barriers
makes a significant suppression of the fission cross section above
5~MeV. At this moment we don't have a simple solution of matching the
kink point in the experimental data and theoretical calculation.

\section{Conclusion}
The Hauser-Feshbach Fission Fragment Decay (HF$^3$D) model was
extended to calculate $\beta$-delayed quantities such as the
cumulative yield calculation and the delayed neutron yield $\nud$,
where consistency of prompt products retained. The model parameters
for $^{235}$U, $^{239}$Pu, and $^{238}$U --- the Gaussian functions to
characterize the primary fission yields, the anisothermal parameter
$R_T$, the spin parameter $f_J$, TKE, and the odd-even term of Wahl's
$Z_p$ model --- were estimated by employing the Bayesian technique
with the KALMAN code at the thermal energy for $^{235}$U, $^{239}$Pu
and 1.2 MeV for $^{238}$U. The result implies that a stronger odd-even
effect is required to reproduce the experimental $\nud$, which is also
reported by Minato~\cite{Minato2018}.

Anchoring the statistical decay calculations to experimental data
available at the thermal energy for $^{235}$U, $^{239}$Pu and 1.2 MeV
for $^{238}$U, we extrapolated the HF$^3$D model to the second chance
fission threshold energy, and demonstrated that the calculated
cumulative FPYs fairly reproduced the experimental data, as well as
$\nup$ and $\nud$ simultaneously. The flat behavior of $\nud$ along
the neutron-incident energy seen in the experimental data of $^{235}$U
and $^{239}$Pu was attributed to a coincidental compensation of
increasing and decreasing delayed neutron precursors.

To examine the sudden change in $\nud$ near 5~MeV, we extrapolated our
calculations beyond the second-chance fission by assuming the same
parameters as the first chance. Indeed this is a crude assumption,
nevertheless we were able to reproduce the step-function-like
variation of $\nud$.  This is promising, and our HF$^3$D model for the
independent and cumulative FPY should be the most advanced tool for
evaluating the FPY data, because it produces many fission observable
quantities in a consistent manner. Unfortunately our calculation drops at
around 5.5~MeV, despite the kink in the experimental data is seen near
4~MeV. This discrepancy should be explained by further investigation
in both the theory and experimental data. Having said that, the
HF$^3$D model qualitatively explains that the variation seen in $\nud$
is a result of different precursors produced by fission at each
fission-chance.

\section*{Acknowledgements}
We thank Dr. Minato for valuable discussions on the delayed neutron
emission calculation.  TK thanks P. Talou, M.B. Chadwick, T. Bredeweg,
and M. Gooden of LANL and A. Tonchev of LLNL for encouraging and
continuous support of this work.  TK and AL performed this work under
the auspice of the U.S. Department of Energy by Los Alamos National
Laboratory under Contract 89233218CNA000001.


\end{document}